\documentclass[10 pt, letterpaper]{article}

\usepackage{amssymb}
\usepackage{amsmath}
\usepackage{IEEEtrantools}
\usepackage{times}
\usepackage[T1]{fontenc}
\usepackage[pdftex]{color,graphicx}
\usepackage{multicol}
\usepackage{multirow}
\usepackage{ctable}
\usepackage[binary-units=true]{siunitx}
\usepackage[ruled,lined,linesnumbered]{algorithm2e} 
\usepackage{etoolbox}			            

\newcommand{\figref}[1]{Fig.~\ref{#1}}
\newcommand{\perc}[1]{\SI{#1}{\percent}}
\newcommand{\degr}[1]{\SI{#1}{\degree}}
\newcommand{\radi}[1]{\SI{#1}{\radian}}
\newcommand{\Meter}[1]{\SI{#1}{\meter}}
\newcommand{\sqm}[1]{\SI{#1}{\meter\squared}}
\newcommand{\mps}[1]{\SI{#1}{\meter\per\second}}
\newcommand{\bits}[1]{\SI{#1}{\bit}}
\newcommand{\Newton}[1]{\SI{#1}{\newton}}
\newcommand{\Hz}[1]{\SI{#1}{\hertz}}
\newcommand{\seconds}[1]{\SI{#1}{\second}}
\newcommand{\kg}[1]{\SI{#1}{\kilo\gram}}

\newlength \figwidth
\newlength \figwidthsmall
\if@twocolumn
\else
\fi

\makeatletter
\patchcmd{\@algocf@start}{%
  \begin{lrbox}{\algocf@algobox}%
}{%
  \rule{0.2\textwidth}{\z@}%
  \begin{lrbox}{\algocf@algobox}%
  \begin{minipage}{0.6\textwidth}%
}{}{}
\patchcmd{\@algocf@finish}{%
  \end{lrbox}%
}{%
  \end{minipage}%
  \end{lrbox}%
}{}{}
\makeatother

\begin{document}
\date{}

\title{Real-time Optimization and Adaptation of the\\ Crosswind Flight of Tethered Wings\\ for Airborne Wind Energy
\thanks{This manuscript is a preprint of a paper submitted for possible publication on the IEEE Transactions on Control Systems Technology and is subject to IEEE Copyright. If accepted, the copy of
record will be available at \textbf{IEEEXplore} library: http://ieeexplore.ieee.org/.}
\thanks{This research has received funding from the California Energy Commission under the EISG grant n. 56983A/10-15 ``Autonomous flexible wings for high-altitude wind energy generation'', from the European Union Seventh Framework Programme (FP7/2007-2013) under grant agreement n. PIOF-GA-2009-252284 - Marie Curie project ``Innovative Control, Identification and Estimation Methodologies for Sustainable Energy Technologies'', and from the Swiss Competence Center Energy and Mobility (CCEM). The authors acknowledge SpeedGoat$^\circledR$'s Greengoat program.}}
\author{A. U. Zgraggen\thanks{Corresponding author: zgraggen@control.ee.ethz.ch.},
 L. Fagiano and M. Morari
\thanks{A. Zgraggen, L. Fagiano and M. Morari are with the Automatic Control Laboratory, Swiss Federal Institute of Technology, Zurich, Switzerland.}}
\maketitle

\begin{abstract}
Airborne wind energy systems aim to generate renewable energy by means of the aerodynamic lift produced by a wing tethered to the ground and controlled to fly crosswind paths. The problem of maximizing the average power developed by the generator, in presence of limited information on wind speed and direction, is considered. At constant tether speed operation, the power is related to the traction force generated by the wing. First, a study of the traction force is presented for a general path parametrization. In particular, the sensitivity of the traction force on the path parameters is analyzed. Then, the results of this analysis are exploited to design an algorithm to maximize the force, hence the power, in real-time. The algorithm uses only the measured traction force on the tether and it is able to adapt the system's operation to maximize the average force with uncertain and time-varying wind. The influence of inaccurate sensor readings and turbulent wind are also discussed. The presented algorithm is not dependent on a specific hardware setup and can act as an extension of existing control structures. Both numerical simulations and experimental results are presented to highlight the effectiveness of the approach.

\end{abstract}

\section{Introduction}\label{sec:intro}
Airborne wind energy (AWE) systems \cite{FaMi12} aim to harness wind energy beyond the altitude of traditional wind mills, in stronger and more steady winds, using tethered wings. The tethers are used to transfer the energy down to the ground. In particular, depending on the system layout, the traction force applied by the wing on the tethers is used to drive generators on the ground, or the energy from on-board generators is transferred via an electrified tether to the ground unit. To increase the power output, the wings are controlled to fly roughly perpendicular to the wind direction \cite{Loyd80}, in so-called crosswind paths. In the recent past, an increasing number of research groups in academia and industry started to develop new concepts of AWE systems, see e.g. \cite{Makani,SkySails,Ampyx,Windlift,Kitenergy,Enerkite,SwissKitePower,CaFM07,IlHD07,CaFM09c,TBSO11,VeGK12}.
The automatic control of the tethered wings plays a major role for the efficiency and thus also economics of such energy generators. The goal is to control the wing in order to fly a crosswind path under constraints such as actuator or wing position limitations, while maximizing the generated power. In order to maximize the power output, the wing should fly on a path that yields the highest traction force for the given wind condition. This problem has been studied by several research groups, see \cite{IlHD07,BaOc12,WiLO08,HouDi07,CoBo13,PhDMoritzDiehl,PhDLorenzoFagiano}. Most of these approaches employ an optimal path, computed off-line for specific wind conditions based on a nonlinear point-mass model.
An automatic controller is then designed to follow this optimal reference trajectory.
Yet, the offline generated optimal trajectories are subject to model-plant mismatch, hence they may be sub-optimal or even infeasible in practice. Moreover, when on-line optimization is used, like in Model Predictive Control approaches, the solution of a complex nonlinear optimization problems is required in real-time, which can be difficult and unreliable. Finally, the mentioned approaches assume that the wind speed and direction at the wing's location are known in order to employ the computed optimal path. However, the wind field changes over distance and time and it is difficult to estimate with only a few measurement points, like those available with ground anemometers.

In order to tackle these issues, in this paper we propose a model-free optimization approach, for the real-time adaptation of the flown paths, assuming no exact knowledge of the wind conditions.
At first, we analyze the influence of the crosswind path on the traction force, in order to asses the most important aspects of the flown trajectory for the sake of power generation. The results indicate that the location of the path with respect to the wind direction and vertical profile has much greater importance than its shape.
Then, we introduce a real-time optimization algorithm aimed to improve and adapt the location, rather than the shape, of a given flown crosswind path using only the measurements of the wing's position relative to the ground and of the traction forces, i.e. no knowledge of the wind direction or profile. 
Additionally, we investigate the effects of erroneous sensor readings and turbulent wind on the performance of the adaptation, showing that the first do not affect the algorithm but the latter can slow down the convergence.
We present the promising results obtained by applying the approach in numerical simulations as well as in real-world experiments.

The paper is organized as follows. We explain the system under consideration and elaborate the problem formulation in Section \ref{sec:system}. Then, we present a study of the traction force of a tethered wing for a flown path in Section \ref{sec:CWTF}. 
Based on this analysis, the proposed algorithm to maximize the traction force is described in Section \ref{sec:algo}.
In Section \ref{sec:measErrAndTurb} the influence of sensor noise and wind turbulences on the algorithm are discussed. In Section \ref{sec:results} numerical simulations and results from tests flights with a small scale prototype are presented.

\section{System description and problem formulation}
\label{sec:system}
We consider an AWE generator that exploits aerodynamic lift to produce electrical energy. For an overview of such systems, see e.g. \cite{FaMi12}. The main components of the generator are the ground unit, the tether, and the wing. The tether is used to anchor the wing to the ground unit, where realizations with one or multiple tethers are possible. The wing is flown on a periodic path, sustained by the aerodynamic lift, which results in a traction force $F$ on the tether. The electricity is either generated on-board of the wing, with small propellers and on-board generators \cite{Makani}, or in the ground unit, by unreeling the tether from drums connected to generators \cite{Ampyx,Windlift,Kitenergy}.  

\begin{figure}[!tbhp]
 \begin{center}
  \includegraphics[trim= 0cm 0cm 0cm 0cm,width=\figwidth]{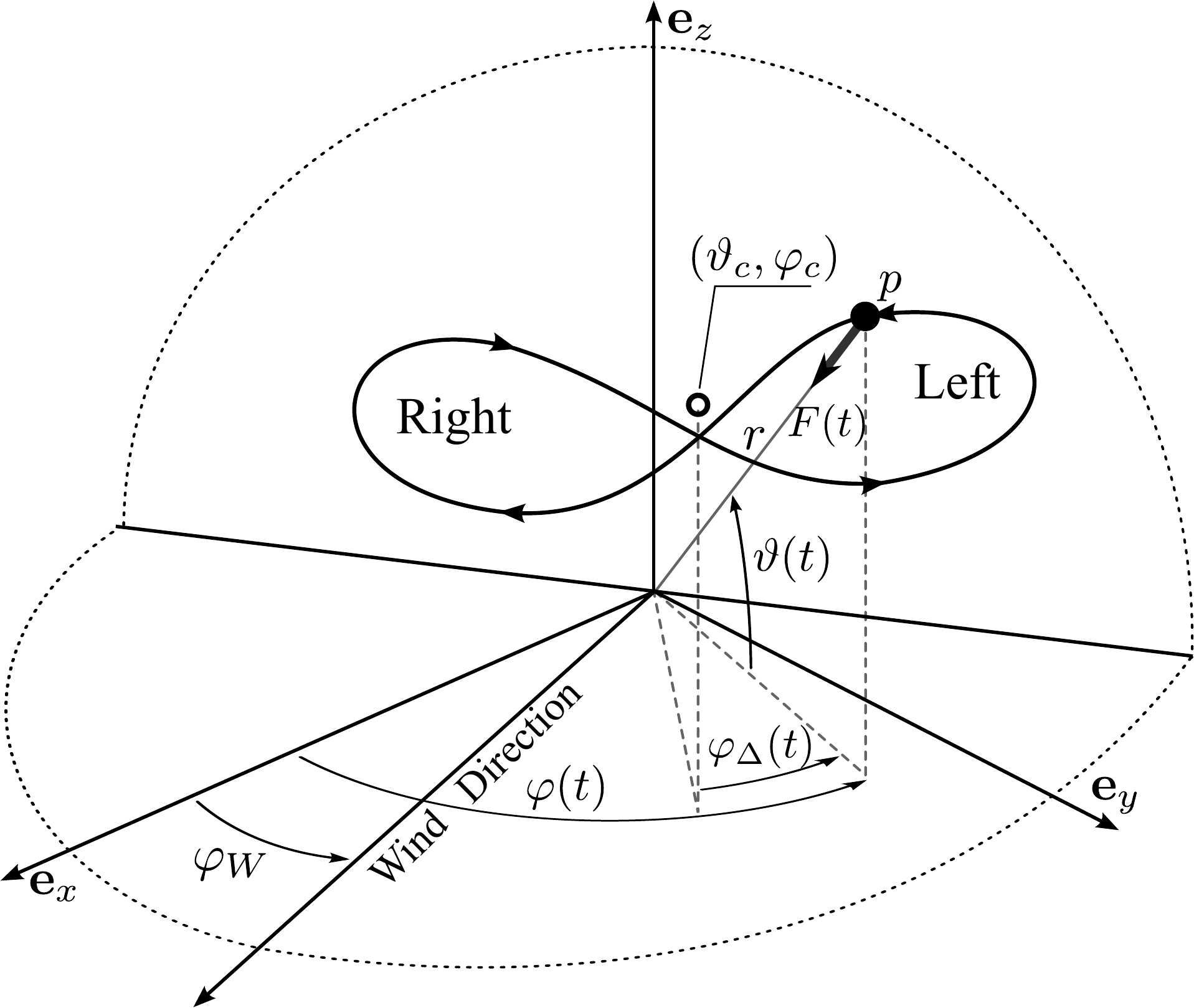}
  \caption[System Layout]{The wing's position $p$ (black dot) is shown on a figure eight path. Note the arrows on the path, indicating an "up-loop" flight pattern (i.e. the wing climbs up on the side of the path and dives in the middle). The wind window is depicted with dotted lines. The average location of the path (circle) is denoted by $(\phi_c,\theta_c)$. The prevalent wind direction forms an angle $\phi_W$ with the fixed axis $\mathbf{e}_x$. }
  \label{fig:ModelCoordFrame}
 \end{center}
\end{figure}

We define a right-handed inertial coordinate system $(\mathbf{e}_x,\mathbf{e}_y,\mathbf{e}_z)$, fixed to the ground unit (see \figref{fig:ModelCoordFrame}). The unit vectors $\mathbf{e}_x$ and $\mathbf{e}_y$ are parallel to the ground and $\mathbf{e}_z$ is vertical with respect to the ground and pointing upwards. The wing's position $p$ is described by spherical coordinates consisting of the two angles $\phi$ and $\theta$ and the tether length $r$. Assuming a straight tether, the azimuthal angle $\phi$ defines the angle between the projection of the tether on the ground and the $\mathbf{e}_x$ axis, while the elevation $\theta$ represents the angle between the tether and the ground plane $(\mathbf{e}_x,\mathbf{e}_y)$.
We assume that the incoming wind is parallel to the ground, i.e. the $(\mathbf{e}_x,\mathbf{e}_y)$-plane, and its misalignment with respect to $\mathbf{e}_x$ is denoted by $\phi_W$, see \figref{fig:ModelCoordFrame}.

Due to boundary layer flow effects of the wind above the earth's surface, the magnitude of the wind $W$ is a function of the altitude $z$ above the ground, the so-called wind shear effect. Common choices to model such a wind shear are the log or the power laws \cite{Ray2006}. In this paper, we consider the latter, but the results hold for a general monotonically increasing wind profile. In our coordinate system, the altitude is given by $z=r\sin\theta$ and the power law is defined as

\begin{figure}[!tbhp]
 \begin{center}
  \includegraphics[trim= 0cm 0cm 0cm 0cm,angle=0,width=\figwidth]{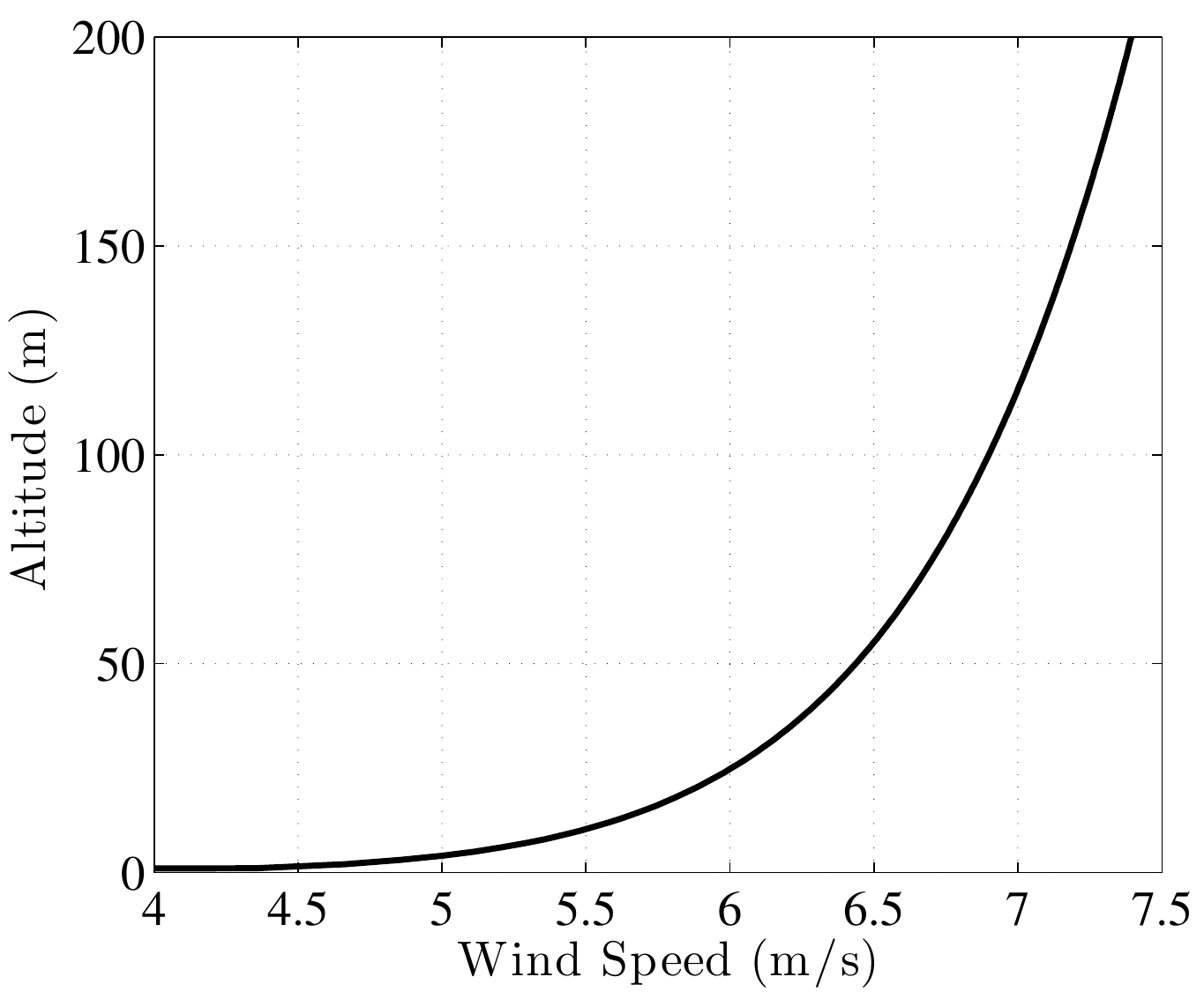}
  \caption[Wind Profile]{Wind profile defined by the power law with $W_0=\mps{5}$, $Z_0=\Meter{4}$, and $\alpha=0.1$}
  \label{fig:PowerLawWindProfile}
 \end{center}
\end{figure}

\begin{IEEEeqnarray}{rCl}\label{eqn:PowerLawWindShear}
 W(\theta) &=& W_0\left(\frac{r \sin\theta}{Z_0}\right)^\alpha\, ,
\end{IEEEeqnarray}
where $W_0$ is the reference wind speed at the reference altitude $Z_0$ and $\alpha$ is the power law exponent, which depends on the roughness of the surface \cite{Ray2006}. In \figref{fig:PowerLawWindProfile} an example of such a wind profile is given.

In AWE systems, during power generation the tethered wing cannot fly upwind, surpassing its anchor point against the wind. Thus, its motion is restricted on a quarter sphere defined by the tether length $r$, the ground plane $(\mathbf{e}_x,\mathbf{e}_y)$, and a vertical plane perpendicular to the prevalent direction of the wind field and containing the anchor point of the tether (see \figref{fig:ModelCoordFrame}, dotted lines). This quarter sphere is called  ``wind window''.
The wing is assumed to fly periodic paths in the wind window, under the action of a feedback controller $K$. 
Such a path can be described by a set of points in the $(\phi,\theta)$-plane. The average position of the path is denoted by $(\phi_c,\theta_c)$. The angular distances from such an average position to each point on the path in $\phi$ and $\theta$ directions are denoted by $\phi_\Delta$ and $\theta_\Delta$, respectively. By introducing the continuous time variable $t$, we can define the corresponding trajectory as the pair
\begin{IEEEeqnarray*}{rClrCl}
 \phi(t)   &=& \phi_c+\phi_\Delta(t)\,,\qquad
 \theta(t) &=& \theta_c+\theta_\Delta(t)
\end{IEEEeqnarray*}
with the trajectory period $T$ to complete one closed path, i.e.
\begin{IEEEeqnarray*}{rClrCl}
 \phi_\Delta(t+T) &=& \phi_\Delta(t)\,,\qquad
 \theta_\Delta(t+T) &=& \theta_\Delta(t)\,.
\end{IEEEeqnarray*}
We define the left and right half paths as the points where $\phi_\Delta(t)\geq 0$ and $\phi_\Delta(t) < 0$, respectively.

\begin{figure}[!tbhp]
 \begin{center}
  \includegraphics[trim= 0cm 0cm 0cm 0cm,width=\figwidthsmall]{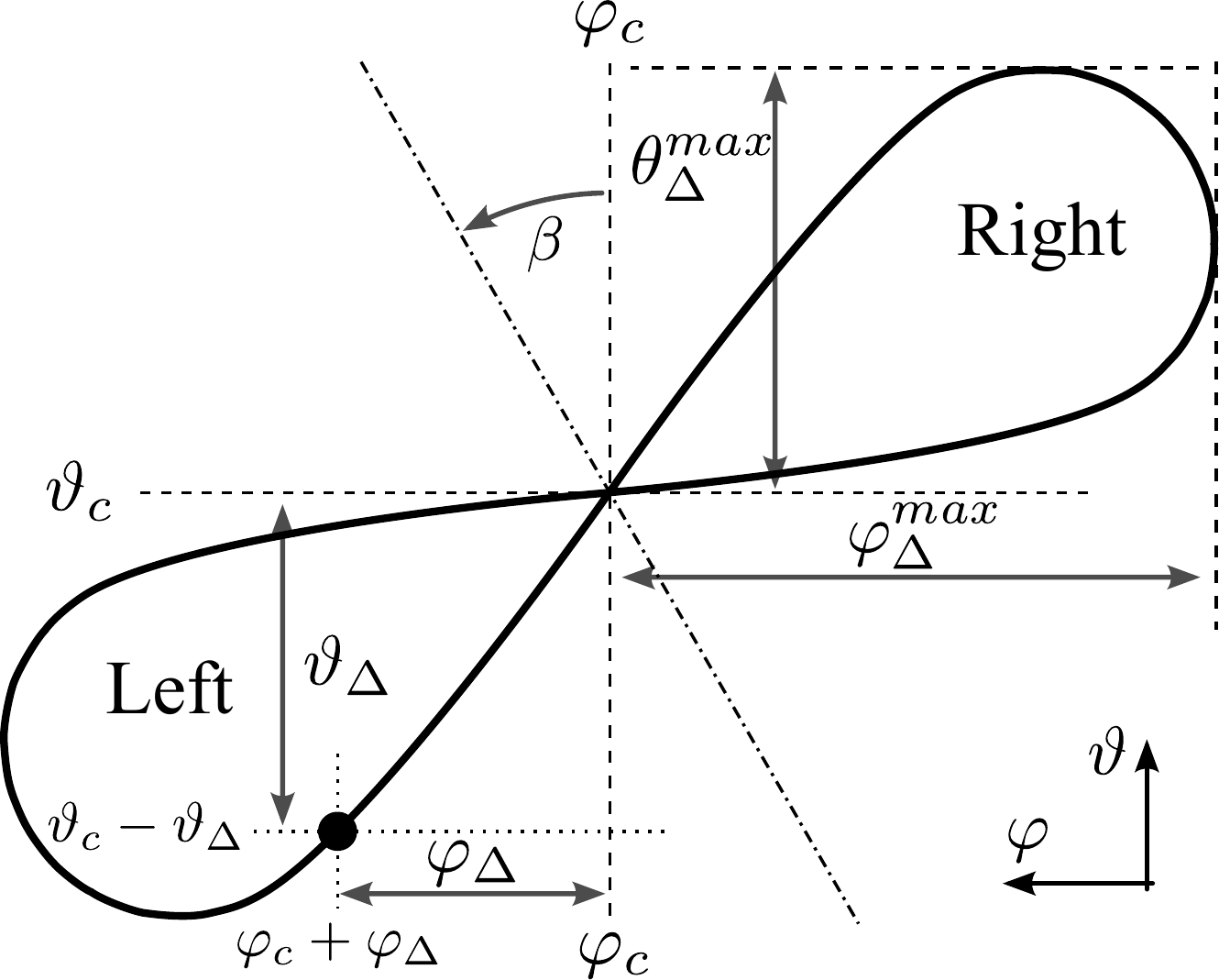}
  \caption[Inclined path]{A generic inclined path with average position $(\phi_c,\theta_c)$ plotted in the $(\phi$-$\theta)$-plane. The $\phi$ and $\theta$ coordinates are depicted as seen from the ground unit and looking at the wing, note the orientation of the $\phi$ axis. A point on the left half path is shown as a black dot. The angle $\beta$ defines the inclination of the path, whose symmetry line is shown as dash-dotted line.}
  \label{fig:GeneralLoopRotated}
 \end{center}
\end{figure}

For systems with multiple tethers, the path has to be such that the tethers do not coil up during one full period and therefore we will consider paths shaped like an eight, see e.g. \cite{IlHD07}, flying up-loops. This means the wing flies upwards on the side and down in the center of the figure eight, see \figref{fig:ModelCoordFrame}. For the remainder of this paper we will call one closed path a ''loop'' or ''path'' to refer to a single flown figure eight. We assume that the path is symmetric w.r.t. a line in the $(\phi,\theta)$-plane. The angle between this symmetry line and the line $\phi=\phi_c$ is denoted with $\beta$, named the ``inclination'' of the path.
The range of $\phi_\Delta$ values is $[-\phi_\Delta^{max},\phi_\Delta^{max}]$ with $\phi_\Delta^{max}>0$. The maximal $\phi_\Delta(t)$ value, $\phi_\Delta^{max}$, defines the lateral span of the path, since it accounts for half of the total lateral span. Similarly, the range of $\theta_\Delta$ values is $[-\theta_\Delta^{max},\theta_\Delta^{max}]$ with $\theta_\Delta^{max}>0$. The maximal $\theta_\Delta(t)$ value, $\theta_\Delta^{max}$, defines the vertical span of the path. 
See \figref{fig:ModelCoordFrame} and \ref{fig:GeneralLoopRotated} for a graphical representation.

The dynamics of the system can be generally described as
\begin{IEEEeqnarray}{rCl}
 \label{eqn:GeneralSystem}
 \dot{x}  &=& f(x,u,\phi_W,W_0,Z_0,\alpha)\\
       y  &=& g(x,u)\, ,\nonumber
\end{IEEEeqnarray}
where $x$ denotes the states, $u$ the control input, and $y$ the measured output. The wind cannot be easily measured or estimated, hence we assume that the wind direction $\phi_W$ and parameters $W_0$, $Z_0$, $\alpha$  are not precisely known. The control input $u$ is computed by the controller $K$, which is a discrete-time system with internal state $z$, input $y$, and parameters $\Theta$. 

\small
\begin{IEEEeqnarray*}{rCl}
 K\:&:& \left\{\begin{matrix}z(\tau +1) &=& h_z(z(\tau),y(\tau),\Theta(\tau))&\\
			     u(t)       &=& h_u(z(\tau),y(\tau),\Theta(\tau)),&\forall t\in[\tau T_s,(\tau +1)T_s)
		\end{matrix}\right. \, ,
\end{IEEEeqnarray*}
\normalsize
where $\tau\in\mathbb{N}$ is the discrete sampling instant and $T_s$ the sampling time.
The parameters $\Theta$ contain specifications of the path to be flown by the wing, namely the average position of the path $(\phi_c,\theta_c)$, its spans $\phi_\Delta^{max}$ and $\theta_\Delta^{max}$, and inclination $\beta$:
\begin{IEEEeqnarray}{rCl}
 \Theta &\doteq& (\phi_c,\theta_c,\phi_\Delta^{max},\theta_\Delta^{max},\beta)\, ,\label{eqn:ThetaParams}
\end{IEEEeqnarray}
It is assumed that the controller $K$ is able to attain such specifications.

The average power $\bar{P}$ produced by a AWE generator with generators on the ground during one full path with period $T$ is
\begin{IEEEeqnarray*}{rCl}
 \bar{P} &=& \frac{1}{T} \int_{0}^{T}{\dot{r}(t)\:F(t)}dt\,,
\end{IEEEeqnarray*}
where $\dot{r}$ is the reel-out speed of the tether and $F(t)$ is the traction force at time $t$. As it is done in several previous work, see e.g. \cite{Loyd80,CaFM09c}, we consider power production at a constant reel-out speed.
Hence, we obtain:
\begin{IEEEeqnarray*}{rCl}
 \bar{P} = \dot{r}\:\bar{F}\, ,
\end{IEEEeqnarray*}
where 
\begin{IEEEeqnarray}{rCl}
 \label{eqn:avgFcontinuous}
 \bar{F} &=& \frac{1}{T}\int_{0}^{T}{F(t)}dt\,.
\end{IEEEeqnarray}

Thus, in this framework the maximization of the average traction force implies maximization of the average power produced during the path. This also holds for systems with on-board generation where turbines
are installed on the wing, since the obtained apparent wind speed is directly related to the traction force \cite{Loyd80}. 

In the considered settings, $\bar{F}$ is a function of the controller's parameters $\Theta$, i.e. our decision variables, and of the wind field, described here by its direction $\phi_W$ and wind shear parameters $W_0$, $Z_0$, $\alpha$, which are uncertain. Our aim is to find the parameters $\Theta$ such that the average traction force (hence the average power) is maximal. 
The related optimization problem can be formulated as
\begin{IEEEeqnarray}{rCl}\label{eqn:GeneralOptProb}
 \underset{\Theta}{\text{max}} &\qquad& \bar{F}(\Theta,\phi_W,W_0,Z_0,\alpha)\, .
\end{IEEEeqnarray}

The exact solution of \eqref{eqn:GeneralOptProb} would require the precise knowledge of the wind profile and direction, which are not assumed to be available here.
In order to tackle this problem, we proceed in two steps. At first, we analyze the influence of $\Theta$ on the average traction force; then, on the basis of such analysis we derive a real-time optimization/adaptation algorithm, to be used on top of controller $K$, able to solve \eqref{eqn:GeneralOptProb} by dealing with the uncertainty of the wind direction and profile, exploiting the measure of the traction force acting on the tethers. For simplicity, in the following we assume the tether length $r$ to be constant, which is a special case of a constant reel-out speed.

\section{Sensitivity Analysis of Crosswind traction force}
\label{sec:CWTF}
In this section, we will first investigate the properties of the average traction force for a flown path using a simplified model. The advantage of such a model is that it allows one to carry out an analytical study of the traction force as a function of the parameters $\Theta$. The results of this first analysis are then compared to simulations of a dynamical non-linear point-mass model of the system.
The latter is derived from first principle equations and includes effects from gravity and inertial forces.

\subsection{Analysis of the traction force with a simplified model}
\label{sec:SCWTFmodel}
A simplified model to estimate the traction force of a tethered wing depending on its location has been introduced in \cite{Loyd80} and subsequently refined in several contributions, for a detailed derivation see e.g. \cite{Loyd80,FaMP11}. According to this model, for a constant reeling speed and fixed values of $W_0,Z_0,\alpha$, the traction force $F$ is a function of the current location of the wing and of the wind direction:

\begin{IEEEeqnarray}{rCl}\label{eqn:CWTFmodel}
 F\left(\theta,\phi,\phi_W\right) &=& \mathcal{C}\: v(\theta)\: m(\phi-\phi_W)\, ,
\end{IEEEeqnarray}
where 
\begin{IEEEeqnarray}{rCl}\label{eqn:CWTFmodelTerms}
  \mathcal{C}    &=& \frac{1}{2}\rho A C_L E_{eq}^2\left(1+\frac{1}{E_{eq}^2}\right)^{\frac{3}{2}}\\
  v(\theta)      &=&  W(\theta)^2 \cos{(\theta)}^2\nonumber\\
  m(\phi-\phi_W) &=&  \cos{(\phi-\phi_W)}^2\nonumber
\end{IEEEeqnarray}
and 
\begin{IEEEeqnarray}{rCl}\label{eqn:CWTFefficiency}
 E_{eq} &=& \frac{C_L}{C_{D,eq}} = \frac{C_L}{C_D+\frac{C_{D,l}A_l}{4A}}\, .
\end{IEEEeqnarray}
In \eqref{eqn:CWTFmodelTerms}-\eqref{eqn:CWTFefficiency}, the air density is indicated by $\rho$, $A$ is the wing reference area, $C_L$ is the wing's lift coefficient, $C_{D,eq}$ is the equivalent drag coefficient, accounting for the drag of the wing and the added drag by the cable. $C_{D,l}$ is the drag coefficient of the cable and $A_l=n_l\,r\,d_l$ is the cable reference area, where $n_l$ is the number of lines holding the wing, $r$ is the line length, and $d_l$ is the line diameter. 
The values of $C_L$ and $C_D$ generally depend on the angle of attack and its derivative, which influence the aerodynamics of the wing. 
However, these coefficients typically do not change much during energy generation. Hence, we assume them to be constant for simplicity, as considered e.g. in \cite{CaFM07}.
For a given wind field, the simplified model \eqref{eqn:CWTFmodel} provides us with a theoretical value of the traction force as a function of the wing's location. Such a theoretical value is obtained by neglecting all forces except for the aerodynamic ones and the cable tension.

\begin{figure}[!tbhp]
 \begin{center}
  \includegraphics[trim= 0cm 0cm 0cm 0cm,angle=0,width=\figwidth]{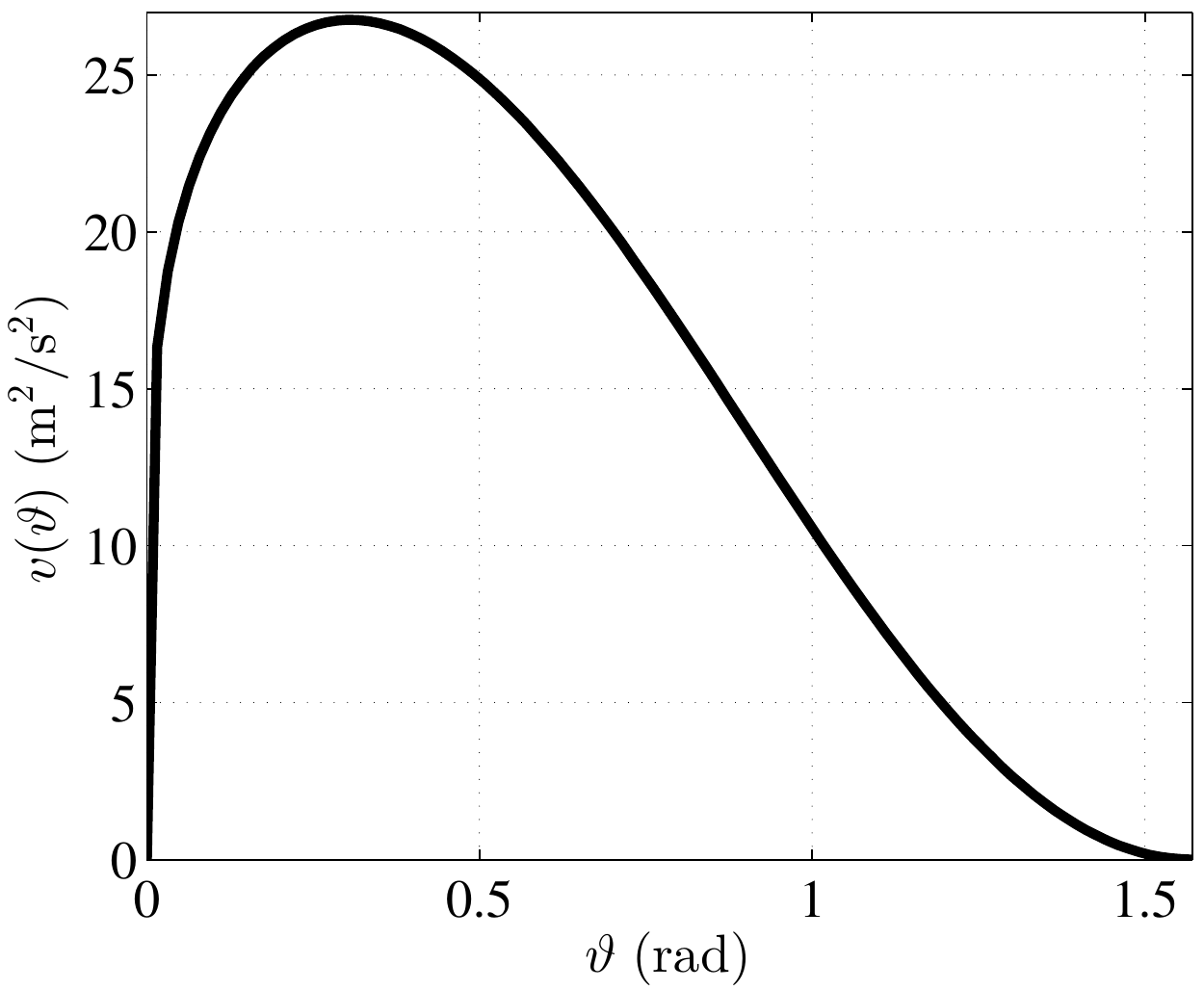}
  \caption{Quasi-concave function $v(\theta)$ with $r=\Meter{30}$, $W_0=\mps{5}$, $Z_0=\Meter{4}$, and $\alpha=0.1$.}
  \label{fig:h_TH_quasiconcave}
 \end{center}
\end{figure}

By inspection, function $m(\phi-\phi_W):(\phi_W-\pi/2,\phi_W+\pi/2)\mapsto(0,1]$ in \eqref{eqn:CWTFmodel}-\eqref{eqn:CWTFefficiency} is quasi-concave with its maximum at $\phi=\phi_W$. Function $v(\theta):(0,\pi/2)\mapsto\mathbb{R^+}$ consists of two parts. The first part, the wind profile $W(\theta)$, is assumed to be monotonically increasing, according to the wind shear model in \eqref{eqn:PowerLawWindShear}, and the second part, $\cos{(\theta)^2}$, is also a quasi-concave function in the domain of $v$. By using the second-order condition for quasi-concave functions \cite{Boyd04}, it can be verified that the product (see \figref{fig:h_TH_quasiconcave} for a typical example) is still quasi-concave and that the point $(\phi,\theta)$ providing maximal traction force for \eqref{eqn:CWTFmodel} is given by $(\phi,\theta)=(\phi_W,\arctan{\left(\sqrt{\alpha}\right)})$.

Equations \eqref{eqn:CWTFmodel}-\eqref{eqn:CWTFefficiency} allow us to carry out an analysis of the traction force as a function of the parameters $\Theta$.
By introducing the index $k=1,\ldots,N$, which identifies the samples of a discretized path with sampling time $T_s$, any sampled position in the path can be expressed as $(\phi_c+\phi_\Delta(k),\theta_c+\theta_\Delta(k))$. 
The discrete form of the average traction force \eqref{eqn:avgFcontinuous} can be then written as
\begin{IEEEeqnarray*}{rCl}
 \frac{1}{T}\int_{0}^{T}{F(t)}dt &\simeq& \frac{1}{NT_s}\sum_{k=1}^{N}{F(k)T_s} = \frac{1}{N}\sum_{k=1}^{N}{F(k)}\, .
\end{IEEEeqnarray*}
The average traction force $\bar{F}$ for one period of the path is thus given by
\small
\begin{IEEEeqnarray}{rCl}\label{eqn:avgF_SCWTF}
 \bar{F}(\Theta,\phi_W,) &=& \frac{1}{N}\sum_{k=1}^{N}{\mathcal{C}\: v(\theta(k))\: m(\phi(k)-\phi_W)}\, ,\\
 \textrm{with}           & & \begin{array}{l}\theta(k) = \theta_c+\theta_\Delta(k)\\ \phi(k) = \phi_c+\phi_\Delta(k)
                             \end{array}\nonumber\, .
\end{IEEEeqnarray}
\normalsize

For the following analysis, we focus on the dependence of $\bar{F}$ on $\phi_c$, $\theta_c$, $\phi_\Delta^{max}$, and $\theta_\Delta^{max}$ only, and fix the inclination $\beta=0$ (see \figref{fig:GeneralLoopRotated}). 
Since the wing is assumed to fly within the wind window, we limit the analysis to the following ranges:
\begin{IEEEeqnarray*}{rCl}\label{eqn:OptProbRanges}
 \phi_c              &\in& (\phi_W-\pi/2,\phi_W+\pi/2)\\
 \theta_c            &\in& (0,\pi/2)\\
 \phi_\Delta^{max}   &\in& (0,\pi/2-(\phi_c-\phi_W)]\\
 \theta_\Delta^{max} &\in& (0,\min{(\theta_c,\pi/2-\theta_c)}]
\end{IEEEeqnarray*}

\begin{figure}[!tbhp]
 \begin{center}
  \includegraphics[trim= 0cm 0cm 0cm 0cm,angle=-90,width=\figwidth]{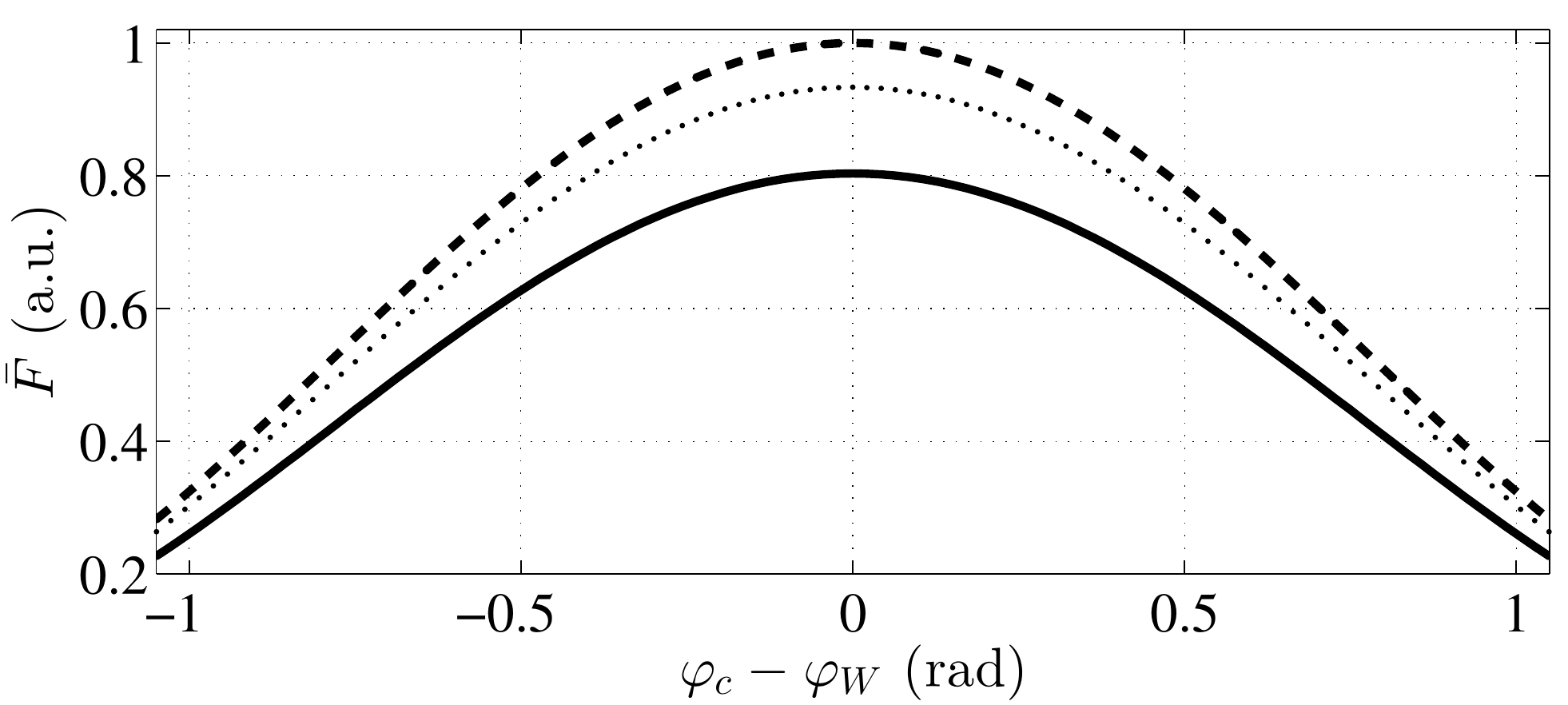}
  \caption{Average traction force computed with the simplified model with the spans of the path $\phi_\Delta^{max}=\radi{0.3}$ and $\theta_\Delta^{max}=\radi{0.1}$. Solid: $\theta_c=\radi{0.1}$, dashed: $\theta_c=\radi{0.3}$, and dotted: $\theta_c=\radi{0.5}$.}
  \label{fig:CWTF_FvsTHc}
 \end{center}
\end{figure}

In \figref{fig:CWTF_FvsTHc}, the average traction force \eqref{eqn:avgF_SCWTF} as a function of $\phi_c-\phi_W$ for three different values of $\theta_c$ is shown. Note that the forces in all the plots have been normalized with the maximum force value of the sample in order to emphasize the independence of the qualitative behavior on the wind: stronger winds influence only the numerical values, but the shape of the curve remains unchanged.
By changing the elevation of the path, $\theta_c$, the value of $\bar{F}$ changes according to the value of $v(\theta)$ from \eqref{eqn:CWTFmodelTerms} (see \figref{fig:CWTF_FvsTHc}). In particular, as it can be inferred by the above-reported discussion on the concavity of the force as a function of $\theta$, there is a single value of $\theta_c$ that maximizes the traction force, and this value depends only on the wind profile and not on the misalignment ($\phi_W-\phi_c$).

From \eqref{eqn:avgF_SCWTF}, we can notice that the contribution of the left and right half-paths to the average traction force $\bar{F}$ are not the same if $\phi_c\neq\phi_W$. We therefore derive the average traction forces for each of the half-paths, and investigate the influence of the parameters $\Theta$ on their difference.
The average traction forces of the left and right half paths are:
\begin{IEEEeqnarray}{rCl}
 \label{eqn:HLavgTF}
 \begin{IEEEeqnarraybox}[][c]{rCl}
  \bar{F}_{L} &=& \frac{1}{N_L}\sum_{k=1}^{N_L}\left\{\mathcal{C}v(\theta(k))m(\phi(k)-\phi_W)|\phi(k)\geq\phi_c\right\}\\
  \bar{F}_{R} &=& \frac{1}{N_R}\sum_{k=1}^{N_R}\left\{\mathcal{C}v(\theta(k))m(\phi(k)-\phi_W)|\phi(k) < \phi_c\right\}\, ,
 \end{IEEEeqnarraybox}
\end{IEEEeqnarray}
where $\bar{F}_{L}$ stands for the average traction force of the left half and $\bar{F}_{R}$ for the right half. $N_L$ and $N_R$ are the number of samples on the left and right half paths, respectively.

The traction force difference between the left and right half-paths, using \eqref{eqn:CWTFmodelTerms} and \eqref{eqn:HLavgTF}, is given, after some manipulations and assuming a sufficiently small sampling time, by

\begin{figure}[!tbhp]
 \begin{center}
  \includegraphics[trim= 0cm 0cm 0cm 0cm,angle=0,width=\figwidthsmall]{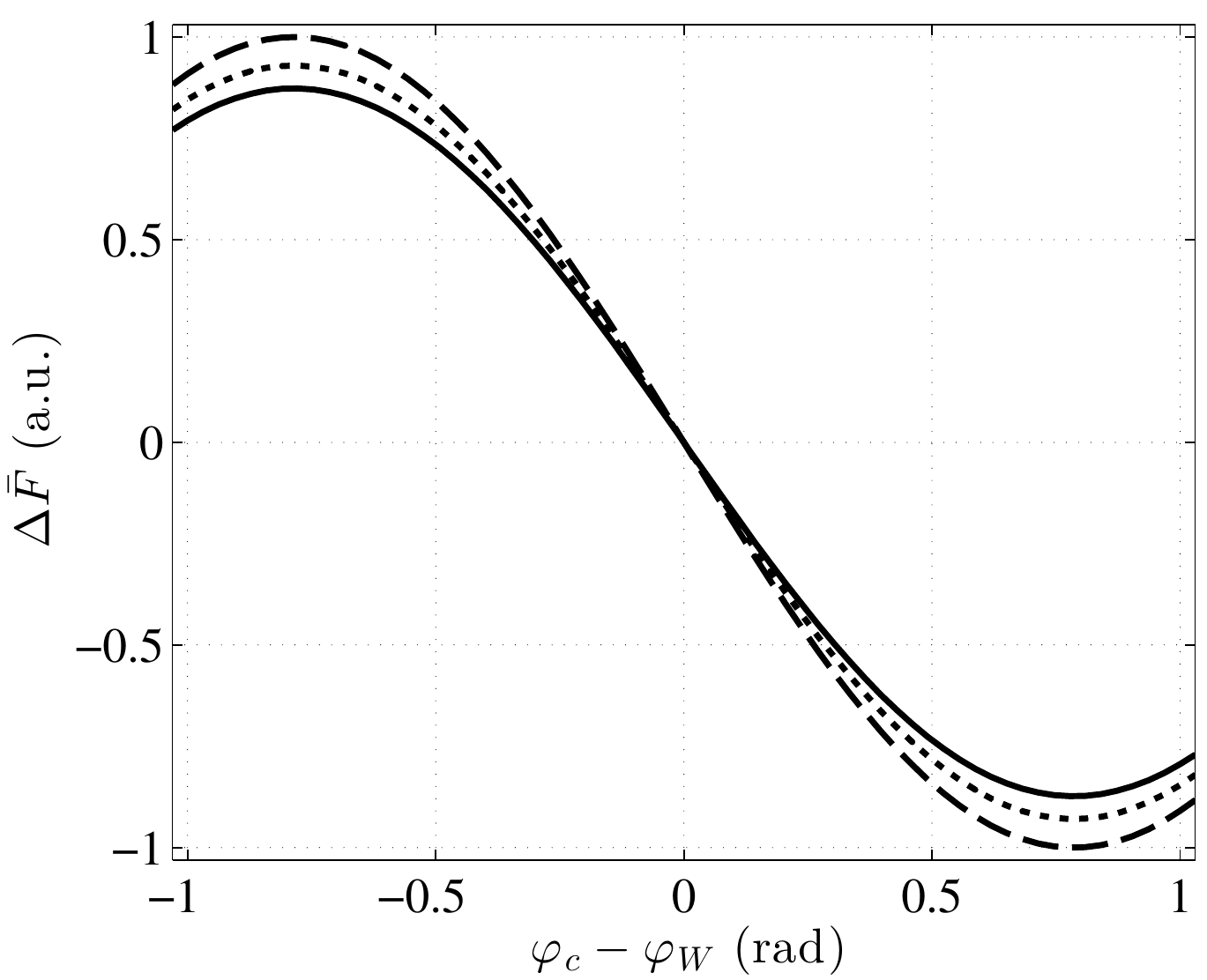}
  \caption{Difference of average traction force $\Delta\bar{F}$ computed with the simplified model, with spans of the path $\phi_\Delta^{max}=\radi{0.3}$ and $\theta_\Delta^{max}=\radi{0.1}$. Solid: $\theta_c=\radi{0.1}$, dashed: $\theta_c=\radi{0.3}$, and dotted: $\theta_c=\radi{0.5}$.}
  \label{fig:CWTF_DFvsTHc}
 \end{center}
\end{figure}

\begin{figure}[!tbhp]
 \begin{center}
  \includegraphics[trim= 0cm 0cm 0cm 0cm,angle=-90,width=\figwidthsmall]{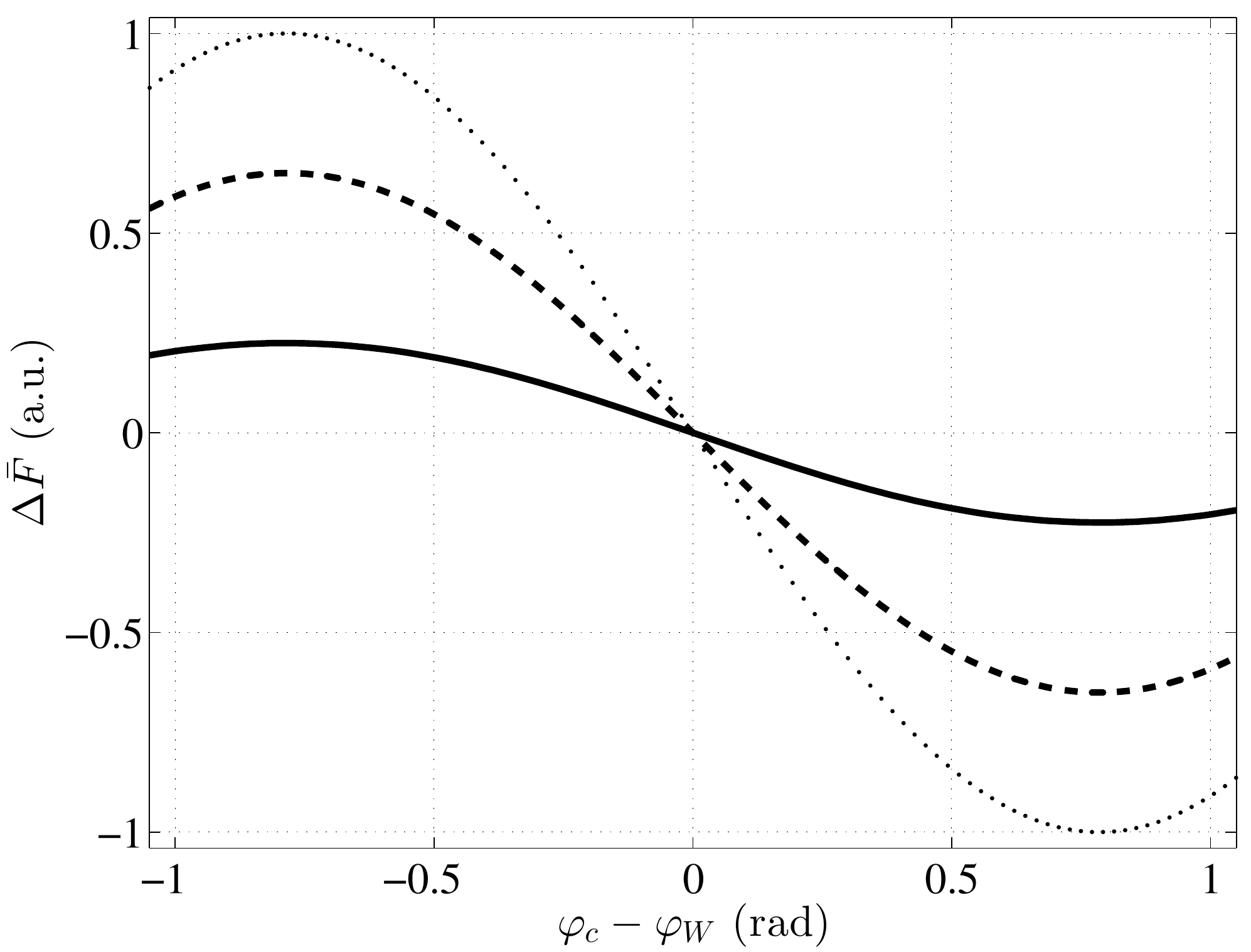}
  \caption{Difference of average traction forces $\Delta\bar{F}$ computed with the simplified model, with $\theta_c=0.2$, $\theta_\Delta^{max}=\radi{0.1}$, and different values of the lateral span. Solid: $\phi_\Delta^{max}=0.1$, dashed: $\phi_\Delta^{max}=0.3$, and dotted: $\phi_\Delta^{max}=0.5$.}
  \label{fig:CWTF_DFvsPHd}
 \end{center}
\end{figure}

\begin{figure}[!tbhp]
 \begin{center}
  \includegraphics[trim= 0cm 0cm 0cm 0cm,angle=-90,width=\figwidth]{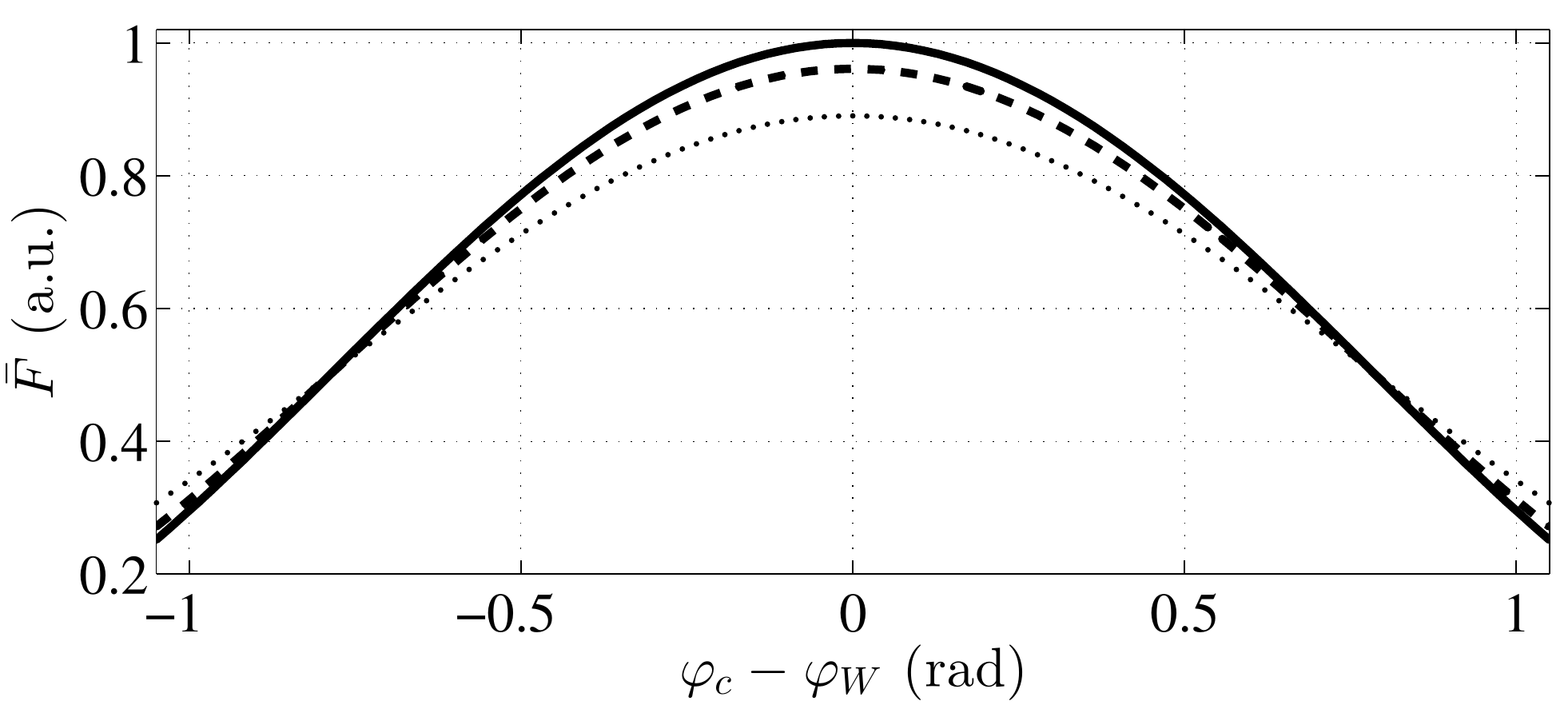}
  \caption{Average traction force $\bar{F}$ computed with the simplified model, with $\theta_c=0.2$, $\theta_\Delta^{max}=\radi{0.1}$, and different values of the lateral span $\phi_\Delta^{max}$. Solid: $\phi_\Delta^{max}=0.1$, dashed: $\phi_\Delta^{max}=0.3$,and dotted: $\phi_\Delta^{max}=0.5$.} 
  \label{fig:CWTF_FvsPHd}
 \end{center}
\end{figure}

\begin{IEEEeqnarray}{rCl}\label{eqn:DF_horizontal}
 \Delta\bar{F}(\Theta,\phi_W) &=& \bar{F}_{L} -\bar{F}_{R} \simeq -\frac{\mathcal{C}}{2}  \sin(2(\phi_c-\phi_W)) \mathcal{B}\,,
\end{IEEEeqnarray}
where the positive term $\mathcal{B}$ is given by
\small
\begin{IEEEeqnarray*}{rCl}
 \mathcal{B} &=& \frac{1}{N_L} \sum_{k=1}^{N_L}{v(\theta(k))\sin(2|\phi_\Delta(k)|)}
                +\frac{1}{N_R} \sum_{k=1}^{N_R}{v(\theta(k))\sin(2|\phi_\Delta(k)|)}
\end{IEEEeqnarray*}
\normalsize
From \eqref{eqn:DF_horizontal} it can be seen that the difference in traction force is zero only if $\phi_c=\phi_W$, i.e. the path is centered w.r.t. the wind, and that it is monotonic for $|\phi_c-\phi_W|\leq\pi/4$. Moreover, paths with an average position on the left of the wind direction, as seen from the anchor point of the tether (i.e. $\phi_c-\phi_W>0$), have a negative $\Delta\bar{F}$, and vice-versa, see \figref{fig:CWTF_DFvsTHc} where the left-right difference in average traction force \eqref{eqn:DF_horizontal} as a function of $\phi_c-\phi_W$ for different values of $\theta_c$ is shown. This comes from the fact that the half-path farther away from the wind direction experiences a smaller fraction of the incoming wind in tether direction, thus generating less traction force.
In \figref{fig:CWTF_DFvsPHd}, a plot of $\Delta\bar{F}$ for different values of the half-span $\phi_\Delta^{max}$ is shown.
By changing the span of the path, the magnitude of $\Delta\bar{F}$ changes. For larger spans, the difference between the average traction force given by the left and right half-paths gets larger, since the average wind conditions for the two halves differ more. The lateral span of the path has also an influence on the average traction force, see \figref{fig:CWTF_FvsPHd}, i.e. wider paths provide smaller average traction force. Thus, a path which has a higher traction force due to its small span will also have a smaller magnitude in $\Delta\bar{F}$ (compare Figs.~\ref{fig:CWTF_DFvsPHd} and \ref{fig:CWTF_FvsPHd}).
Note that the value of $\theta_c$ has an effect on the average traction force difference, too, but this is not as large as that of the span of the path in $\phi$ direction (compare Figs.~\ref{fig:CWTF_DFvsTHc} and \ref{fig:CWTF_DFvsPHd}).

\begin{figure}[!tbhp]
 \begin{center}
  \includegraphics[trim= 0cm 0cm 0cm 0cm,angle=0,width=\figwidthsmall]{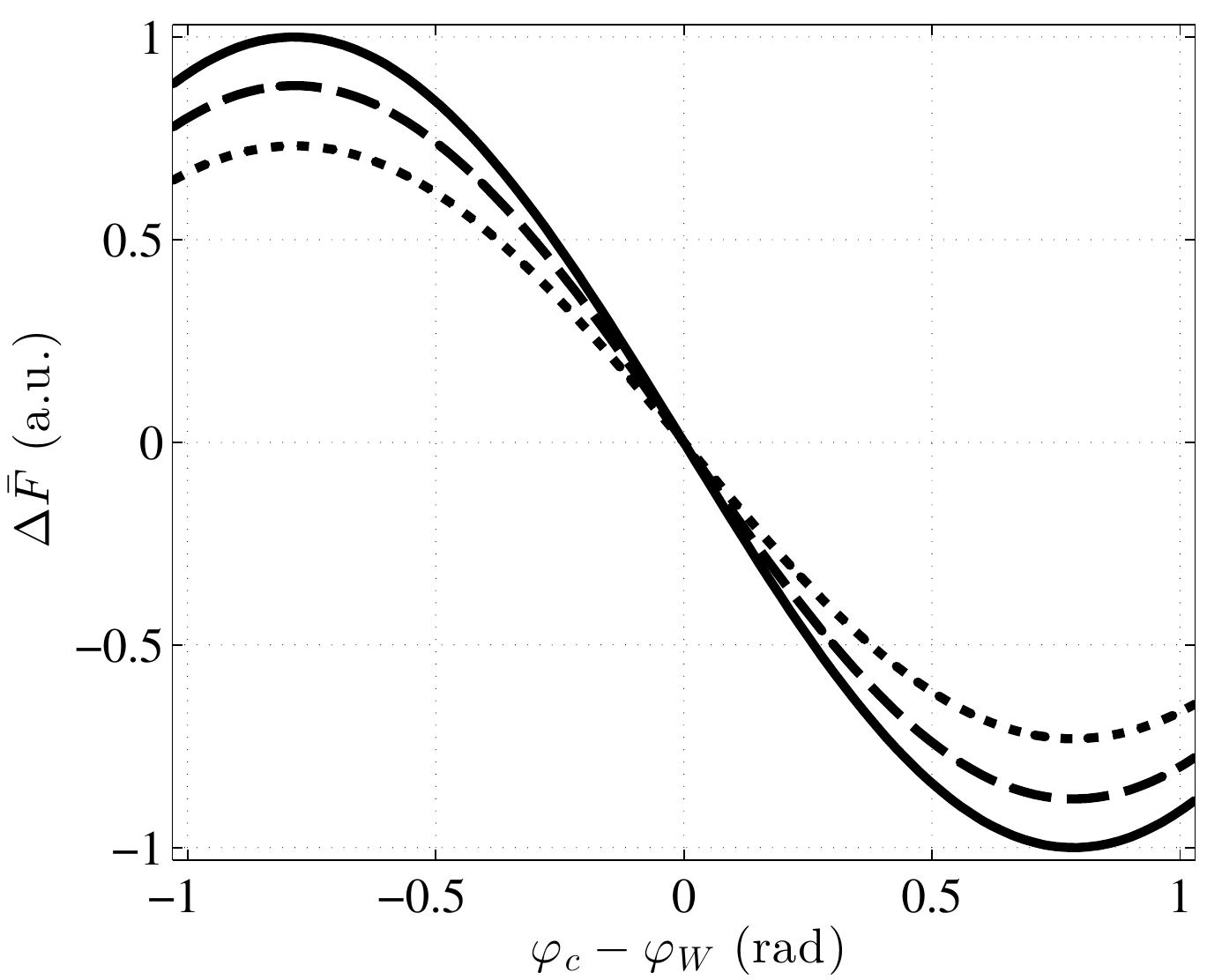}
  \caption{Difference of average traction force $\Delta\bar{F}$ computed with the simplified model, with $\theta_c=0.2$ and a lateral span of the path $\phi_\Delta^{max}=\radi{0.3}$. Solid: $\theta_\Delta^{max}=\radi{0.1}$, dashed: $\theta_\Delta^{max}=\radi{0.3}$, and dotted: $\theta_\Delta^{max}=\radi{0.5}$.}
  \label{fig:CWTF_DFvsTHd}
 \end{center}
\end{figure}

\begin{figure}[!tbhp]
 \begin{center}
  \includegraphics[trim= 0cm 0cm 0cm 0cm,angle=0,width=\figwidth]{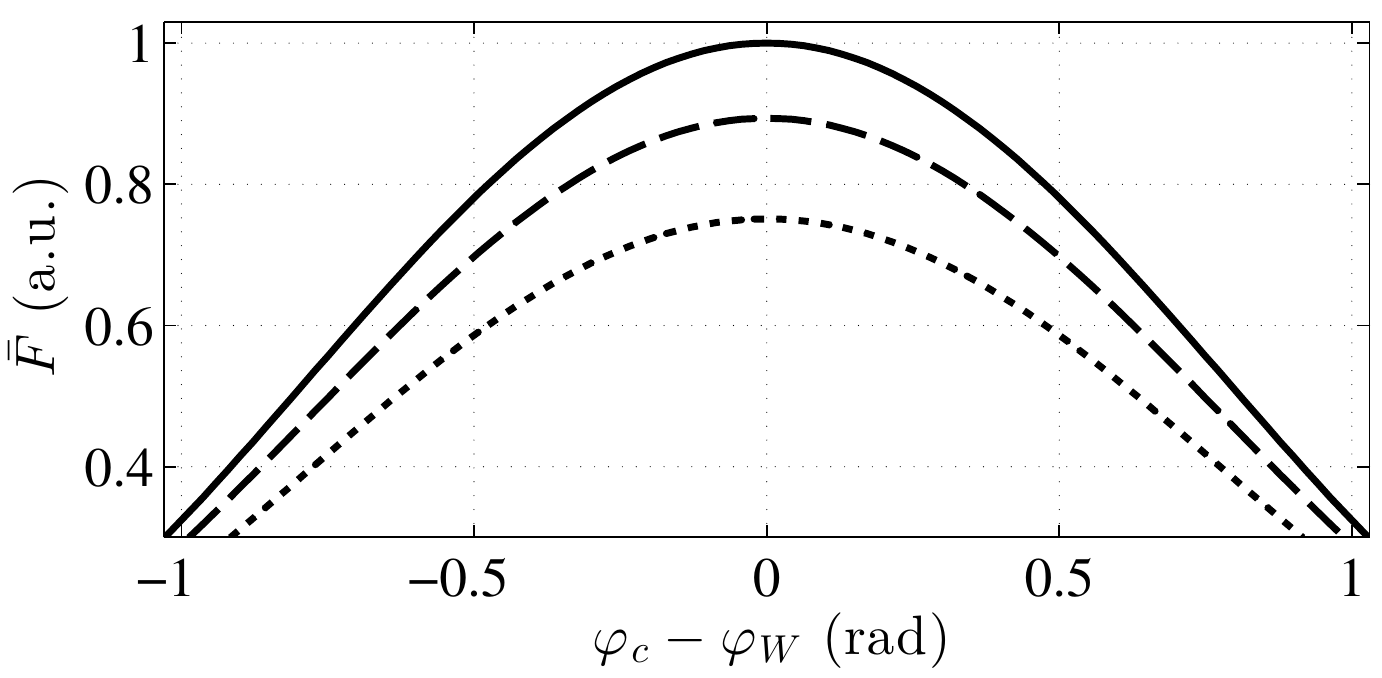}
  \caption{Average traction force $\bar{F}$ computed with the simplified model, with $\theta_c=0.2$ and a lateral span of the path $\phi_\Delta^{max}=\radi{0.3}$. Solid: $\theta_\Delta^{max}=\radi{0.1}$, dashed: $\theta_\Delta^{max}=\radi{0.3}$, and dotted: $\theta_\Delta^{max}=\radi{0.5}$.}
  \label{fig:CWTF_FvsTHd}
 \end{center}
\end{figure}

The span $\theta_\Delta^{max}$ also has an influence on the average traction force $\bar{F}$ and on the difference of left-right average traction forces $\Delta\bar{F}$ as shown in \figref{fig:CWTF_DFvsTHd} and \figref{fig:CWTF_FvsTHd}, respectively. 
Comparing Figs.~\ref{fig:CWTF_DFvsPHd}-\ref{fig:CWTF_FvsTHd}, it can be seen that the span $\phi_\Delta^{max}$ has more influence on the difference between left and right average traction forces whereas $\theta_\Delta^{max}$ has more influence on the total average traction force.

\subsection{Analysis of the traction force with a dynamic model}
\label{sec:PMmodel}
In this section, we employ a dynamic model to asses, via numerical simulations, the considerations derived with the simplified model, and to analyze also the effects of different path inclinations $\beta$.
The dynamics $f(x,u,\phi_W)$ are modeled here by the widely used nonlinear point-mass model for a tethered wing, see e.g. \cite{CaFM07,IlHD07,CaFM09c,WiLO08,PhDMoritzDiehl,PhDLorenzoFagiano}.
The dynamic equations are derived from first principles and the wing is assumed to be a point with given mass. The tether is assumed to be straight with a 
non-zero diameter. The aerodynamic drag of the tether and half of the tether mass are added to the wing's drag and mass, respectively. The aerodynamic forces are modeled with constant lift and drag coefficients, and effects from gravity and inertial forces are included. 
The wing is assumed to be steered by a change of the roll angle $\psi$, which is manipulated by a control system, and thus, referring to \eqref{eqn:GeneralSystem}, we have $u=\psi$. The state $x$ of this system is given by $x=(\phi,\theta,r,\dot{\phi},\dot{\theta},\dot{r})$.

In order to carry out the simulations, the controller $K$ is designed using the approach described in \cite{FaZg13}. Such a controller is able to make the wing fly on a symmetric figure eight path with the required spans and inclination, and with the average position being a given reference location $(\phi_c,\theta_c)$.

Consistently with Section \ref{sec:SCWTFmodel}, the average traction forces generated during the full path and the average traction force generated on the left and right half paths are computed from the simulation results:
\begin{IEEEeqnarray}{rCl}\label{eqn:avgF_k}
  \bar{F}(\Theta,\phi_W)     &=& \frac{1}{N}\sum_{k=1}^{N}{F(k)}
\end{IEEEeqnarray}
\begin{IEEEeqnarray}{C}
 \begin{IEEEeqnarraybox}[][c]{rCl}
   \bar{F}_{L}               &=& \frac{1}{N_L}\sum_{k=1}^{N_L}{\left\{F(k)\: | \phi(k)\geq\phi_c \right\}}\\
   \bar{F}_{R}               &=& \frac{1}{N_R}\sum_{k=1}^{N_R}{\left\{F(k)\: | \phi(k) < \phi_c \right\}}
 \end{IEEEeqnarraybox}\label{eqn:HLavgTF_k}
\end{IEEEeqnarray}
The traction force difference between the left and right half path is
\begin{IEEEeqnarray}{rCl}\label{eqn:avgDF_k}
 \Delta\bar{F}(\Theta,\phi_W) &=& \bar{F}_{L} - \bar{F}_{R}\, .
\end{IEEEeqnarray}

As done before, we want to study how the average traction force and the difference in average traction force between the left and right half paths change for different values of $\Theta$, including this time also the inclination $\beta$, in the range $\beta\in [-\pi/2,\pi/2]$.

\begin{figure*}
\centering
\begin{minipage}[b]{\figwidth}
      \centering\footnotesize
       a)
      \includegraphics[trim= 0cm 0cm 0cm 0cm,angle=-90,width=\figwidth]{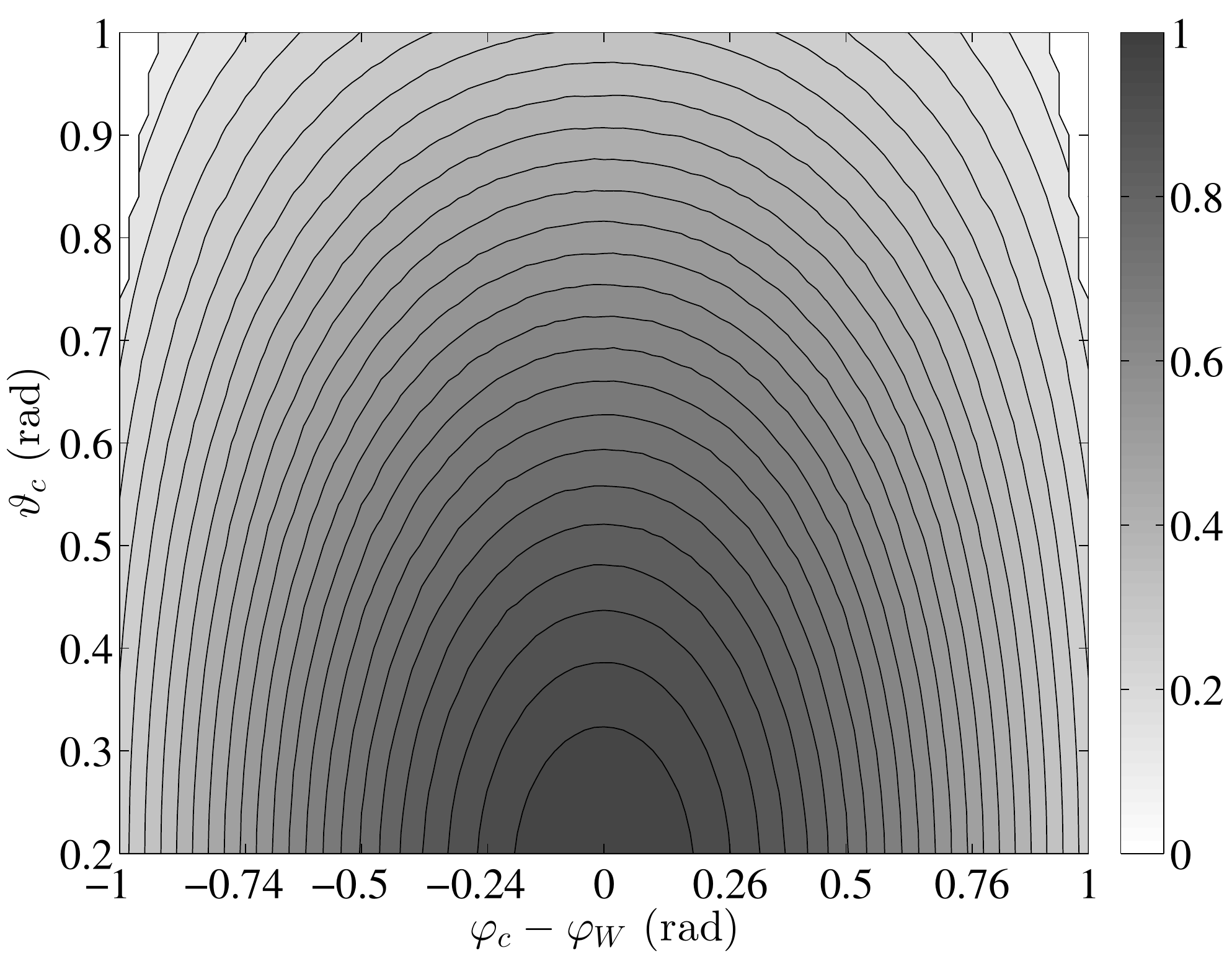}
\end{minipage}\qquad
\begin{minipage}[b]{\figwidth}
      \centering\footnotesize
      b)
      \includegraphics[trim= 0cm 0cm 0cm 0cm,angle=-90,width=\figwidth]{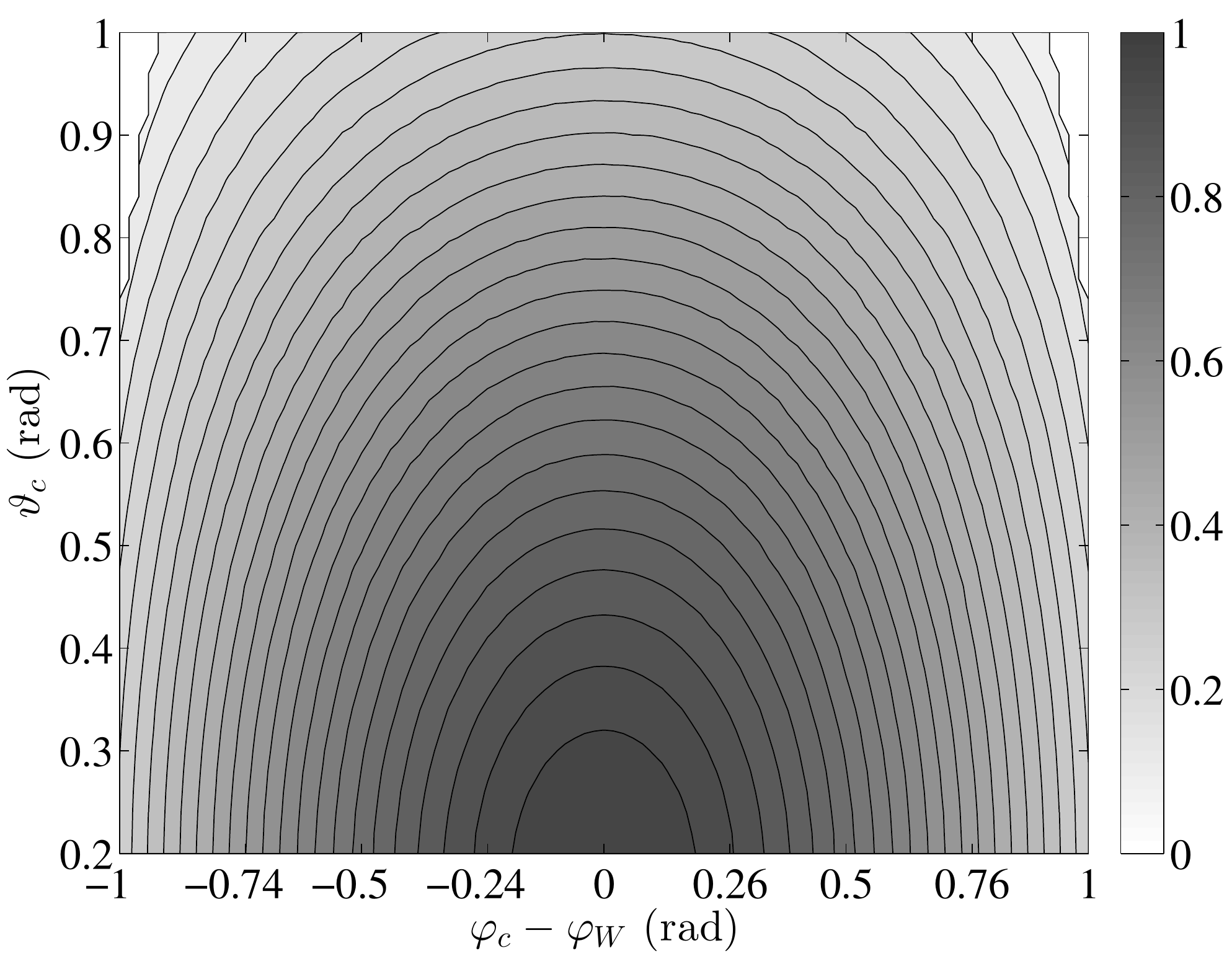}
\end{minipage}
\caption{Contour plots of the average traction force $\bar{F}$, as a function of $\phi_c-\phi_W$ and $\theta_c$, computed with the simplified model a) and with the point-mass model b)}
\label{fig:CWTF_PM_Fcontour}
\end{figure*}

\begin{figure}[!tbhp]
  \begin{center}
    \includegraphics[trim= 0cm 0cm 0cm 0cm,angle=-90,width=\figwidth]{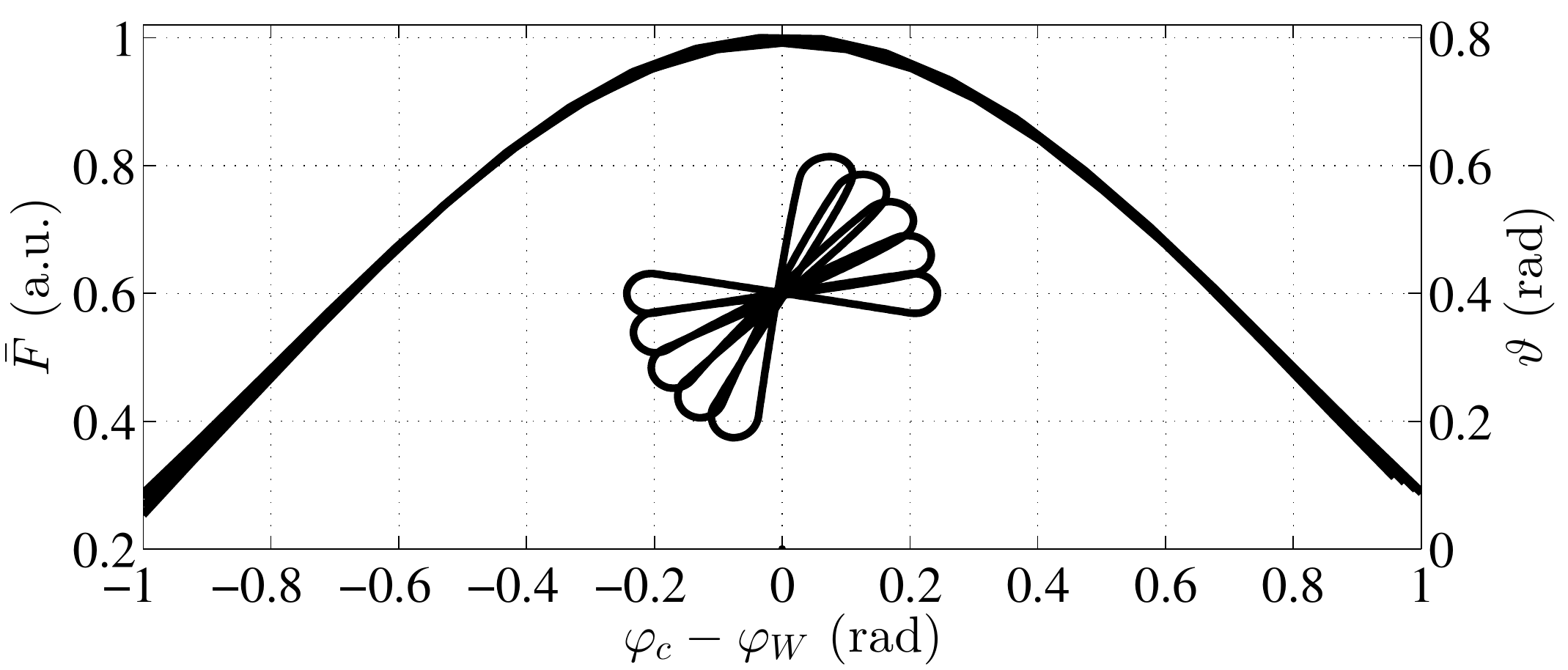}
    \caption{Traction force $\bar{F}$ computed with the point-mass model, as a function of $\phi_c-\phi_W$, with $\theta_c=0.4$. There are five lines with values of $\beta=\{0,0.3,0.6,0.9,1.2\}$~rad. The resulting shapes of the path are depicted underneath the traction force curve; the corresponding y-axis, giving the values of $\theta$ for the flown path, is depicted on the right of the plot.}
    \label{fig:PM_Fvsbeta}
  \end{center}
\end{figure}

\begin{figure}[!htbp]
 \begin{center}
  \includegraphics[trim= 0cm 0cm 0cm 0cm,angle=-90,width=\figwidthsmall]{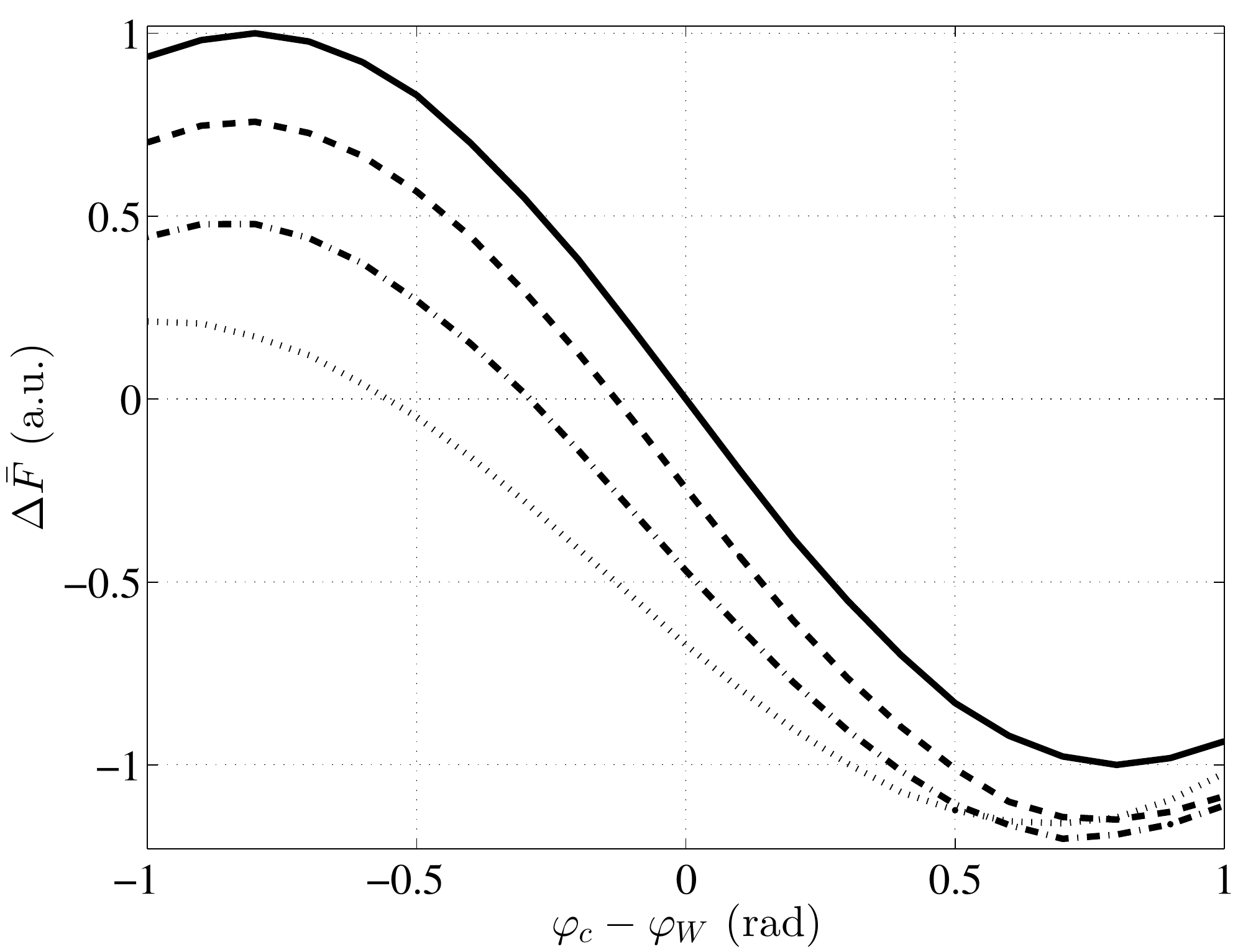}
  \caption[]{Traction force difference $\Delta\bar{F}$ computed with the point-mass model for $\theta_c=0.4$, $\phi_\Delta^{max}=0.24$, and different inclinations $\beta=0$ (solid), $\beta=0.3$ (dashed), $\beta=0.6$ (dot-dashed), and $\beta=0.9$ (dotted).}
  \label{fig:PM_DFvsbeta}
 \end{center}
\end{figure}

Comparing the traction force for various $\phi_c$ and $\theta_c$ with a symmetric horizontal path shape (i.e. $\beta=0$) shows good qualitative correspondence with the simplified model used in Section \ref{sec:SCWTFmodel}, see \figref{fig:CWTF_PM_Fcontour},
thus indicating that gravity and inertial forces do not have impact on the average forces. If the path is inclined, i.e. $\beta\neq0$, the average traction force does not increase more than $\perc{2}$ for $\phi_c$ around the optimum of $\bar{F}$, see \figref{fig:PM_Fvsbeta}, but the values of $\Delta\bar{F}$ change significantly. In fact, when the path is inclined, the traction force difference is not zero anymore for $\phi_c-\phi_W=0$. A positive value of $\beta$ corresponds to a negative value of $\phi_c-\phi_W$ such that $\Delta\bar{F} = 0$ and vice versa, see \figref{fig:PM_DFvsbeta}. The effect of larger spans in the presence of $\beta\neq 0$ is the same as the one observed in Section \ref{sec:SCWTFmodel}, e.g. a larger value of $\phi_\Delta^{max}$ increases $\Delta\bar{F}$ for fixed values of the other parameters. 
As expected from the analysis with the simplified model, stronger wind or different tether length $r$ do not affect the qualitative results.

\subsection{Discussion}
The results of the previous two sections show that there is a single optimal average location, denoted as $(\phi_c^*,\theta_c^*)$, yielding the maximal average traction force for a given path shape. In particular, we have $\phi^*_c=\phi_W$, while $\theta^*_c$ depends on the vertical wind profile. The average traction force is very sensitive on the average position of the path. A misalignment in $\phi_c$ with respect to $\phi_c^*$ of roughly $\degr{20}$ can lead to a decrease of average traction force of $\perc{15}$, while $\perc{50}$ decrease of the force is obtained for a misalignment of roughly $\degr{45}$, see \figref{fig:CWTF_FvsTHc}. An average elevation $\theta_c\neq \theta_c^*$ can also reduce the traction force by a significant amount, in the same order as for $\phi_c$. 
As an example, with an error in both $\phi_c$ and $\theta_c$ of around $\degr{20}$ from the optimum, the traction force will be reduced by almost $\perc{30}$. 

For horizontal paths (i.e. $\beta=0$), the difference in average traction force, $\Delta\bar{F}$, is zero for an average position $\phi_c=\phi_W$ and it is monotonically increasing for values of $\phi_c-\phi_W$ between $\pi/4$ and $-\pi/4$. Moreover, the sign of $\Delta\bar{F}$ is the opposite w.r.t. that of $\phi_c-\phi_W$, i.e. $\phi_c-\phi_c^*$. Therefore, if $\beta=0$ the value of $\Delta\bar{F}$ is a good indicator of the alignment of $\phi_c$ with the wind direction $\phi_W$. As seen in \figref{fig:PM_Fvsbeta}, the inclination has only a small influence on the traction force, changing it less than $\perc{2}$ in the interval around the optimum. However, the average traction force difference between the left and right half paths is sensitive to changes in $\beta$. A positive value of $\beta$ can decrease the magnitude of $\Delta\bar{F}$ on the right side of the wind window up to $\perc{75}$, while increasing it on the left side by only around $\perc{20}$. Thus, with $\beta\neq 0$ the difference in traction force $\Delta\bar{F}$ is not anymore zero for $\phi_c=\phi_W$, and its sign is not the opposite w.r.t the sign of the misalignment $\phi_c-\phi_W$ anymore (see \figref{fig:PM_DFvsbeta}).

As seen in \figref{fig:CWTF_FvsPHd}, increasing the lateral span $\phi_\Delta^{max}$ of the path by a factor of $5$ decreases the average traction force roughly by $\perc{10}$. A larger lateral span of the path scales up $\Delta\bar{F}$, but also decreases the average traction force $\bar{F}$. This could lead to the conclusion that the shorter the lateral span, the better it is in terms of system operation. This is only partially true. First, a short span implies sharp turns that induce more drag, slowing the wing down, hence decreasing the average traction force. However, this effect is not captured by the point-mass model considered here, thus leading to the result that paths with very small span do not lose performance in terms of traction force. Second, for a short span it might be difficult to infer something about the wind direction, due to the small value of $\Delta\bar{F}$. 
As seen in \figref{fig:CWTF_FvsTHd}, increasing the vertical span $\theta_\Delta^{max}$ of the path by a factor of $5$ decreases the average traction force roughly by $\perc{15}$ and also decreases the magnitude of $\Delta\bar{F}$. This shows that a small vertical span $\theta_\Delta^{max}$ of the path is favorable. This has two advantages. First, the wing does not need to overcome gravity to climb for a long distance and secondly it will stay closer to the targeted $\theta_c$ position. 

In conclusion, the analysis above shows that optimizing the average position $(\phi_c,\theta_c)$ yields the largest increase of average traction force (hence generated power). The shape of the path, in terms of lateral span and inclination, has only a relatively small influence on the traction force.
Moreover, even an optimal path, in terms of shape, has to be flown at the optimal location in order not to lose a large fraction of the traction force. 
In the next section, we exploit these considerations to derive an algorithm able to optimize in real-time the average path location and to adapt it in the presence of changing wind direction $\phi_W$, using only the measurements of traction force on the tethers.

\section{Real-time optimization and adaptation algorithm}
\label{sec:algo}
As seen in the previous section, the average location of a flown path has the largest influence on the generated traction force among all of the considered parameters. Thus we aim to find the best average location in $\phi$ and $\theta$ for a given path shape, in order to maximize the average power output of the AWE system. Since the inclination of a path has an adverse effect on $\Delta\bar{F}$ and does not affect $\bar{F}$ much, we only consider horizontal paths with $\beta=0$. Moreover, an inclined path can increase the difference between the maximal and minimal instantaneous traction force $F(t)$ between the left and right half-loops, leading to an asymmetric wear of the system's components. Enforcing a horizontal path can be done in practice with a suitable controller as in \cite{FaZg13}. Motivated by these results, in the following we take only $(\phi_c,\theta_c)$ as free optimization variables out of the considered parameters $\Theta$ in \eqref{eqn:ThetaParams}, while we fix the half-span $\phi_\Delta^{max}$  and the vertical span $\theta_\Delta^{max}$ to prescribed values and select $\beta=0$.

Recall that we assume that the underlying controller $K$ accepts reference values for the average location where the path should be flown. The algorithm we present next will then compute such reference values in order to solve the following optimization problem:

\begin{IEEEeqnarray}{rCl}
 \label{eqn:OptProb}
 \underset{\theta_c,\phi_c}{\text{max}} &\qquad& \bar{F}(\phi_c,\theta_c,\phi_W,W_0,Z_0,\alpha)\,.
\end{IEEEeqnarray}

We assume that the parameters $\phi_W,W_0,Z_0,\alpha$, specifying the wind direction and profile, are not known, hence the optimization problem \eqref{eqn:OptProb} is uncertain due to the lack of information on $\phi_W$ and the wind shear profile. On the other hand, we assume that the traction force $F$ is measured, as well as the position of the wing w.r.t. the ground unit.  Hence the values of $\bar{F}$ and $\Delta\bar{F}$ for each flown path are measured.

The analysis presented in the previous section indicates that we can reformulate the optimization problem \eqref{eqn:OptProb} as
\begin{IEEEeqnarray}{rCl}\label{eqn:2StepOptProb}
 \underset{\theta_c}{\text{max}}\: \left[\underset{\phi_c}{\text{max}} \quad \bar{F}(\phi_c,\theta_c,\phi_W,W_0,Z_0,\alpha)\right]\,,
\end{IEEEeqnarray}
i.e. \eqref{eqn:2StepOptProb} can be maximized separately in $\phi_c$ and $\theta_c$, since the value of $\phi_c$ that maximizes the average traction force, for given $\theta_c$, depends only on $\phi_W$ and not on $\theta_c$ itself and, vice-versa, the optimal value $\theta^*_c$ does not depend on $\phi_c$. Also, note that for horizontal paths we have, irrespective of $\theta_c$,

\small
\begin{IEEEeqnarray*}{rCl}
 \text{arg}\,\underset{\phi_c}{\max}\:\bar{F}(\phi_c,\theta_c,\phi_W,W_0,Z_0,\alpha) &=& \text{arg}\,\underset{\phi_c}{\min}\:|\Delta\bar{F}(\phi_c,\theta_c,\phi_W,W_0,Z_0,\alpha)|
\end{IEEEeqnarray*}
\normalsize
as it can be derived from \eqref{eqn:avgF_SCWTF} and \eqref{eqn:DF_horizontal} and from the results in Section \ref{sec:PMmodel}. Therefore, the problem \eqref{eqn:OptProb} can be solved by addressing two subsequent optimization problems independently. We will first exploit the measure of $\Delta\bar{F}$ to find the best location in $\phi$, i.e. to compute $\text{arg}\min_{\phi_c}\:|\Delta\bar{F}(\phi_c,\theta_c,\phi_W,W_0,Z_0,\alpha)|$, and then the measure of $\bar{F}$ to find the best location in $\theta$, i.e. solving \eqref{eqn:2StepOptProb} with the previously found optimal $\phi_c$.
The advantage of using the differences in average traction forces to find the optimal $\phi_c$, instead of using only $\bar{F}$, is that a single value of $\Delta\bar{F}$, i.e. a single flown path, gives already an indication on the sign of the misalignment $\phi_c-\phi_W$, hence on the search direction for $\phi_c$. By using only $\bar{F}$, the values obtained by two paths with different $\phi_c$ would be needed to estimate the search direction, which would take at least twice as long.
Thus, the adaptation in $\phi$ direction is sped up by looking at the traction force difference $\Delta\bar{F}$ instead of the total average traction force $\bar{F}$ only.

\subsection{Algorithm Outline}
We present a short outline of an algorithm able to adapt the average position of a path, such that it converges to the optimum.
The algorithm iterates over subsequent complete paths flown by the wing, and exploits the values of $\bar{F}$ and $\Delta\bar{F}$ measured in the current and past paths. See Algorithm~\ref{algo:BasicMethdod}.
\begin{algorithm}\label{algo:BasicMethdod}
  \caption{Optimization/Adaptation}
  \DontPrintSemicolon
  \While{ true }{
    \If{ one complete loop flown \nllabel{algoLine:firstIF}}{
      calculate $\Delta\bar{F}$ and $\bar{F}$\;
      \eIf{ $|\Delta\bar{F}| > \Delta\bar{F}_{min}$ }{
	$\underset{\phi_c}{\text{min}}\quad |\Delta\bar{F}|$\;
	update $\phi_c$\;
      }{
	$\underset{\theta_c}{\text{max}}\quad \bar{F}$\;
	update $\theta_c$\;
      }
    }
  }
\end{algorithm}
A more detailed algorithm that can be used in practice, and has been tested during real-world experiments, is given in the Appendix.
The algorithm uses a coordinate search approach, see e.g. \cite{DFO09}, to solve the two subsequent optimization problems, since no gradient information is available.
The algorithm first checks if the absolute value of the traction force difference, $\Delta\bar{F}$, is smaller than some margin, $\Delta\bar{F}_{min}$. The latter is used as a stopping criterion for the $\phi$ direction adaptation. If this condition is not met, the Algorithm adapts the value of $\phi_c$ in order to reduce the absolute value of the force difference, $|\Delta\bar{F}|$. Otherwise, the algorithm searches for the best vertical position $\theta_c$, without changing $\phi_c$. 
Apart from other minor tolerances (see the Appendix for details), the scalar $\Delta\bar{F}_{min}$ is the only tuning parameter in our real-time optimization approach. If $\Delta\bar{F}_{min}$ is very small, the algorithm will tend to spend most its time correcting the azimuthal position of the loop and the $\theta_c$ position would improve slowly. Vice versa, if $\Delta\bar{F}_{min}$ is large, most time is spent in correcting the average elevation position of the flown path.

\section{Effects of Measurement Errors and Turbulence}
\label{sec:measErrAndTurb}
The algorithm introduced in the previous section exploits the measurements of the tether force and wing position, and its performance will clearly depend on the accuracy of the related sensors, as well as on the intensity of wind turbulence. It is therefore of interest to study the effects of such phenomena.

First, we start with the description of the expected sensor errors, followed by an analysis of how much these errors affect the adaptation algorithm. Secondly, we will present the effects of added turbulence on the wind profile, and introduce measures to counteract its effects. We carry out these analyses mainly with the simplified traction force model used in section \ref{sec:SCWTFmodel}. We will also rely on simulation results employing the point-mass model of the wing to highlight specific effects, using a standard turbulence model which is common in wind turbine analysis.

\subsection{Sensor Errors}
For this analysis we assume that line angle sensors, measuring the orientation of the tethers with respect to the ground unit, are used to estimate the position of the wing. Moreover, an on-board inertial measurement unit can additionally be used to improve position data \cite{FHBK12}.
The line angle sensors are assumed to be optical encoders measuring the angle between the tethers and the ground, for the elevation angle $\theta$, and between the projection of the tethers on the ground and the symmetry axis of the ground unit, for the azimuthal angle $\phi$. Such encoders have various sources of errors
and in general the accuracy is usually around $\pm1$ count, e.g an encoder with a \bits{10} resolution has an additive error of less than \degr{0.4}.
 
The additive error from the encoder can be expressed as
\begin{IEEEeqnarray}{rCl}
 \tilde{\phi}   &=& \phi+\epsilon_\phi\label{eqn:PHAngleError}\\
 \tilde{\theta} &=& \theta+\epsilon_\theta\label{eqn:THAngleError}
\end{IEEEeqnarray}
where the variables with "~$\tilde{}~$" denote noise-corrupted measurements, variables without a "~$\tilde{}~$" the true values and $\epsilon$ represents the additive error. 
The estimated center location of the path can be written as
\begin{IEEEeqnarray}{rCl}
 \tilde{\phi}_c   &=& \frac{1}{N}\sum_{k=1}^N{\phi(k)+\epsilon_\phi(k)}\label{eqn:PHcTilde}\\
 \tilde{\theta}_c &=& \frac{1}{N}\sum_{k=1}^N{\theta(k)+\epsilon_\theta(k)}\, .\label{eqn:THcTilde}
\end{IEEEeqnarray}
Again, $k\in[1,N]$ where one loop flown by the wing has $N$ sample points.
The center location of the flown path is derived from all measurement points captured during one loop. 
Assuming that the error at each time step is i.i.d. with zero mean, we can expect that the error on the center location of the path in $\phi$ and $\theta$ converges to zero for high sampling rates.
We can in fact rewrite \eqref{eqn:PHcTilde} and \eqref{eqn:THcTilde} as
\begin{IEEEeqnarray*}{rCl}
 \tilde{\phi}_c   &=& \frac{1}{N}\sum_{k=1}^N{\phi(k)}+\frac{1}{N}\sum_{k=1}^N{\epsilon_\phi(k)} \approx \phi_c + \bar{\epsilon}_\phi = \phi_c \\ 
 \tilde{\theta}_c &=& \frac{1}{N}\sum_{k=1}^N{\theta(k)}+\frac{1}{N}\sum_{k=1}^N{\epsilon_\theta(k)} \approx \theta_c + \bar{\epsilon}_\theta = \theta_c \, . 
\end{IEEEeqnarray*}

The difference in average traction force between the left and right half-paths can be calculated according to \eqref{eqn:DF_horizontal}, using \eqref{eqn:PHAngleError} and \eqref{eqn:THAngleError}; with some manipulations and assuming a sufficiently small sampling time, we obtain
\begin{IEEEeqnarray}{rCl}\label{eqn:DF_horizontal_AngleError}
 \Delta\bar{F}(\Theta,\phi_W) &=& \bar{F}_{L} -\bar{F}_{R} \simeq -\frac{\mathcal{C}}{2}  \sin(2(\phi_c-\phi_W)) \tilde{\mathcal{B}}\, ,
\end{IEEEeqnarray}
where the positive term $\tilde{\mathcal{B}}$ is given by
\begin{IEEEeqnarray*}{rCl}
 \tilde{\mathcal{B}} &=& \frac{1}{N_L} \sum_{k=1}^{N_L}{v(\tilde{\theta}(k))\sin(2(|\tilde{\phi}_\Delta(k)|+\epsilon_\phi(k)))}\\
                     & & +\frac{1}{N_R} \sum_{k=1}^{N_R}{v(\tilde{\theta}(k))\sin(2(|\tilde{\phi}_\Delta(k)|+\epsilon_\phi(k)))}
\end{IEEEeqnarray*}
From \eqref{eqn:DF_horizontal_AngleError} we see again that the difference in average traction force is zero only if $\phi_c=\phi_W$ and that it is monotonic for $|\phi_c-\phi_W|\leq \pi/4$.
From the equation above it can be seen that the influence of the line angle sensors' errors on $\Delta F$ is not changing its qualitative shape, or the estimation of the average path position. Therefore, the alignment in $\phi$ direction is not affected by the line angle sensor errors. 

Similarly, the alignment in $\theta$ direction, using the average traction force $\bar{F}$ \eqref{eqn:avgF_SCWTF}, is not influenced by the errors in the line angle sensors.
Thus, such errors have no impact on the performance of the adaptation algorithm.

We next consider errors affecting the force measurements. Since the tether force is the main feedback variable used by our algorithm, we expect the related errors to be more critical for the performance of our adaptation approach. The force sensors are load cells installed at ground level, see e.g. \cite{FaMa13} for details. The related measurement error is assumed to consist of two components, an additive term and a multiplicative term:
\begin{IEEEeqnarray*}{rCl}
 \tilde{F}(k) &=& F(k)(1+\delta_F)+\epsilon_F(k)
\end{IEEEeqnarray*}
$\delta_F$ accounts for a calibration error of the signal gain of the sensor and $\epsilon_F$ is an additive bias accounting for noise with zero mean plus a calibration offset. It is assumed that $|\delta_F|<1$.
Thus, for a sufficiently small sampling time the effects of the additive term on the average force become constant:
\begin{IEEEeqnarray}{rCl}
 \label{eqn:AddavgTF_Ferror}
 \bar{\epsilon}_F &\approx& \frac{1}{N}\sum_{k=1}^{N}\epsilon_F(k)\, .
\end{IEEEeqnarray}

The difference in average traction force can be written as
\begin{IEEEeqnarray}{rCl}\label{eqn:measDTF}
 \Delta\tilde{\bar{F}} &=& \tilde{\bar{F}}_L-\tilde{\bar{F}}_R
\end{IEEEeqnarray}
where
\small
\begin{IEEEeqnarray}{rCl}
 \label{eqn:HLavgTF_Ferror}
 \begin{IEEEeqnarraybox}[][c]{rCl}
  \tilde{\bar{F}}_{L} &=& \frac{1}{N_L}\sum_{k=1}^{N_L}\left\{\mathcal{C}v(\theta(k))m(\phi(k)-\phi_W)\left(1+\delta_F\right)+\epsilon_F(k)|\phi(k)\geq\phi_c\right\}\\
  \tilde{\bar{F}}_{R} &=& \frac{1}{N_R}\sum_{k=1}^{N_R}\left\{\mathcal{C}v(\theta(k))m(\phi(k)-\phi_W)\left(1+\delta_F\right)+\epsilon_F(k)|\phi(k) < \phi_c\right\}\, .
 \end{IEEEeqnarraybox}
\end{IEEEeqnarray}
\normalsize

\begin{figure}[!htbp]
 \begin{center}
  \includegraphics[trim= 0cm 0cm 0cm 0cm,angle=0,width=\figwidthsmall]{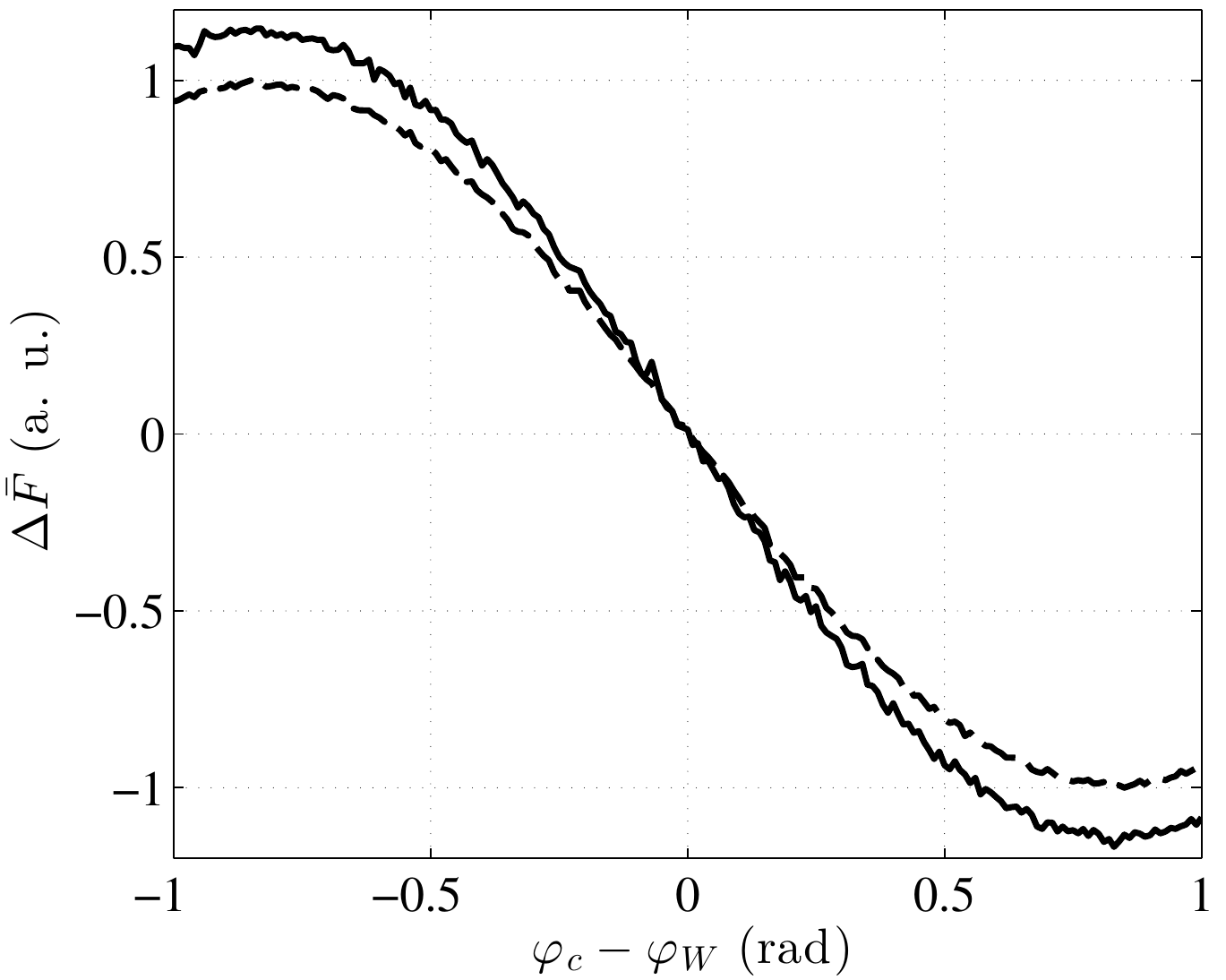}
  \caption[]{Simulation results. Difference in average traction force $\Delta\bar{F}$ as a function of the misalignment between the average loop location and the wind direction, computed using the point-mass model with $\delta_F=0.15$, $\epsilon_F=250\pm100~\Newton{}$ (normally distributed). The true $\Delta\bar{F}$ (dashed) and measured $\Delta\tilde{\bar{F}}$ (solid) are shown.}
  \label{fig:DFvsPHc_FSensErr}
 \end{center}
\end{figure}

Using \eqref{eqn:AddavgTF_Ferror}, Equations \eqref{eqn:measDTF}-\eqref{eqn:HLavgTF_Ferror} can be simplified to
\begin{IEEEeqnarray}{rCl}\label{eqn:DF_horizontal_ForceError}
 \Delta\tilde{\bar{F}}(\Theta,\phi_W) &=& \tilde{\bar{F}}_{L}-\tilde{\bar{F}}_{R} \simeq -\frac{\mathcal{C}}{2}  \sin(2(\phi_c-\phi_W)) \tilde{\mathcal{B}}\, ,
\end{IEEEeqnarray}
where the term $\tilde{\mathcal{B}}$ is given by
\begin{IEEEeqnarray*}{rCl}
 \tilde{\mathcal{B}} &=& \frac{1}{N_L} \sum_{k=1}^{N_L}{v(\theta(k))\sin(2|\phi_\Delta(k)|)(1+\delta_F)}\\
                     & & +\frac{1}{N_R}\sum_{k=1}^{N_R}{v(\theta(k))\sin(2|\phi_\Delta(k)|)(1+\delta_F)}\\
                     & & +\bar{\epsilon}_{F_L}-\bar{\epsilon}_{F_R}
\end{IEEEeqnarray*}
From \eqref{eqn:DF_horizontal_ForceError} we see that the qualitative behavior of $\Delta\tilde{\bar{F}}$ as a function of $\phi$ is the same as that of the true values, hence 
the alignment in $\phi$ direction is not affected by the force sensor errors, see \figref{fig:DFvsPHc_FSensErr} for a simulation example.
It has to be noted that for high sampling rates the values $\bar{\epsilon}_{F_L}$ and $\bar{\epsilon}_{F_R}$ are the same, since the same force sensors are used for both halves of the path, hence their effect in \eqref{eqn:DF_horizontal_ForceError} cancels out.

\begin{figure}[!htbp]
 \begin{center}
  \includegraphics[trim= 0cm 0cm 0cm 0cm,angle=0,width=\figwidthsmall]{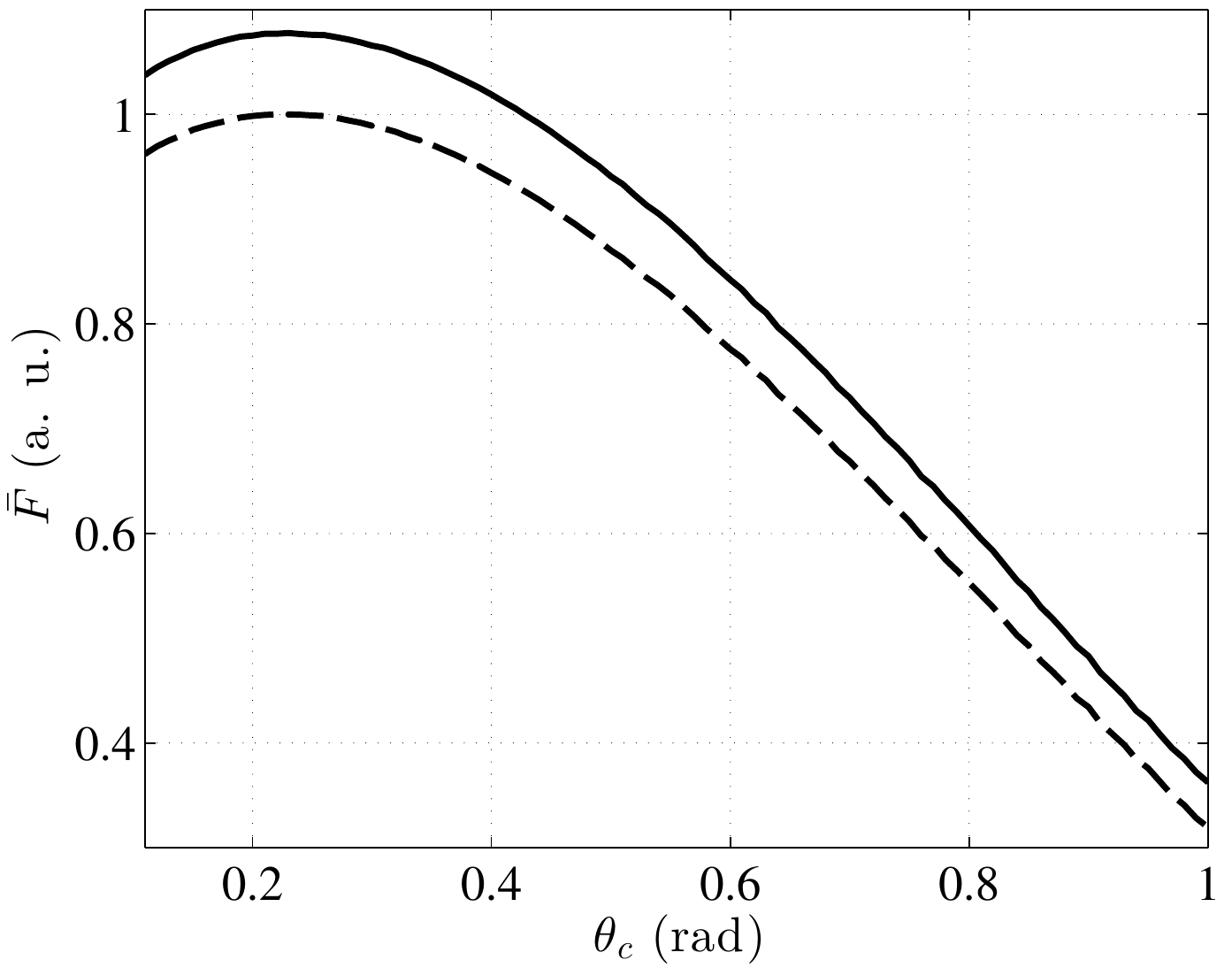}
  \caption[]{Simulation results. Average traction force $\bar{F}$ as a function of the average loop elevation $\theta_c$, computed using the point-mass model with $\delta_F=0.15$, $\epsilon_F=250\pm100~\Newton{}$ (normally distributed). The true $\bar{F}$ (dashed) and the measured $\tilde{\bar{F}}$ (solid) are shown.}
  \label{fig:FvsTHc_FSensErr}
 \end{center}
\end{figure}

For the alignment in $\theta$ direction, we consider the difference of the average traction force of two loops at different elevation angle $\theta_c$, $\bar{F}_1$ and $\bar{F}_2$, since such a difference is used in our approach to compute the search direction for $\theta_c$. Also in this case, it can be shown that the considered force sensor errors do not affect the alignment algorithm. Assume that two loops at different elevation angles $\theta_c$ were flown with average traction forces $\bar{F}_1$ and $\bar{F}_2$. 
Additionally, without loss of generality, we can assume that $\bar{F}_1>\bar{F}_2$. The difference of the \emph{measured} average traction forces can then be written as
\begin{IEEEeqnarray*}{rCl}
 \tilde{\bar{F}}_1-\tilde{\bar{F}}_2 &=&      \bar{F}_1-\bar{F}_2 + \bar{F}_1\delta_F-\bar{F}_2\delta_F+\bar{\epsilon}_F-\bar{\epsilon}_F\\
                                     &\simeq& \left(\bar{F}_1-\bar{F}_2\right)\left(1+\delta_F\right)\, .
\end{IEEEeqnarray*}
From this we can see that the additive error term does not play a role, and that the multiplicative error only changes the magnitude of the difference of the two flown loops, but not the sign (which is used in the optimization/adaptation algorithm), i.e. the qualitative behavior of the measured force as a function of $\vartheta_c$ is the same as that of the true force, see \figref{fig:FvsTHc_FSensErr} for an example.

\subsection{Turbulences}
In a real-world system, the incoming wind will never be perfectly smooth and some fluctuations, such as wind gusts or turbulences, will be present. Using the wind shear profile \eqref{eqn:PowerLawWindShear}, this can be expressed as
\begin{IEEEeqnarray*}{rCl}
 W_t(k,\phi,\theta) &=& W_0\left(\frac{r\sin{(\theta)}}{Z_0}\right)^\alpha + W_\Delta(k,\phi,\theta) 
\end{IEEEeqnarray*}
where $W_t$ stands for the wind profile with added turbulences and $W_\Delta$ is the change in wind speed around the nominal wind value due to turbulences for a given point in time and space. For the sake of simplicity of notation, we omitted the sampling time $k$ for the position angles $\phi$ and $\theta$.
As it can be seen from \eqref{eqn:CWTFmodel}, changes in the wind speed influence the traction force quadratically. 
Thus, $W_\Delta$ will significantly affect the traction force developed by the wing and, consequently, the performance that can be achieved by the adaptation algorithm. 

\begin{figure}[!htbp]
 \begin{center}
  \includegraphics[trim= 0cm 0cm 0cm 0cm,angle=0,width=\figwidth]{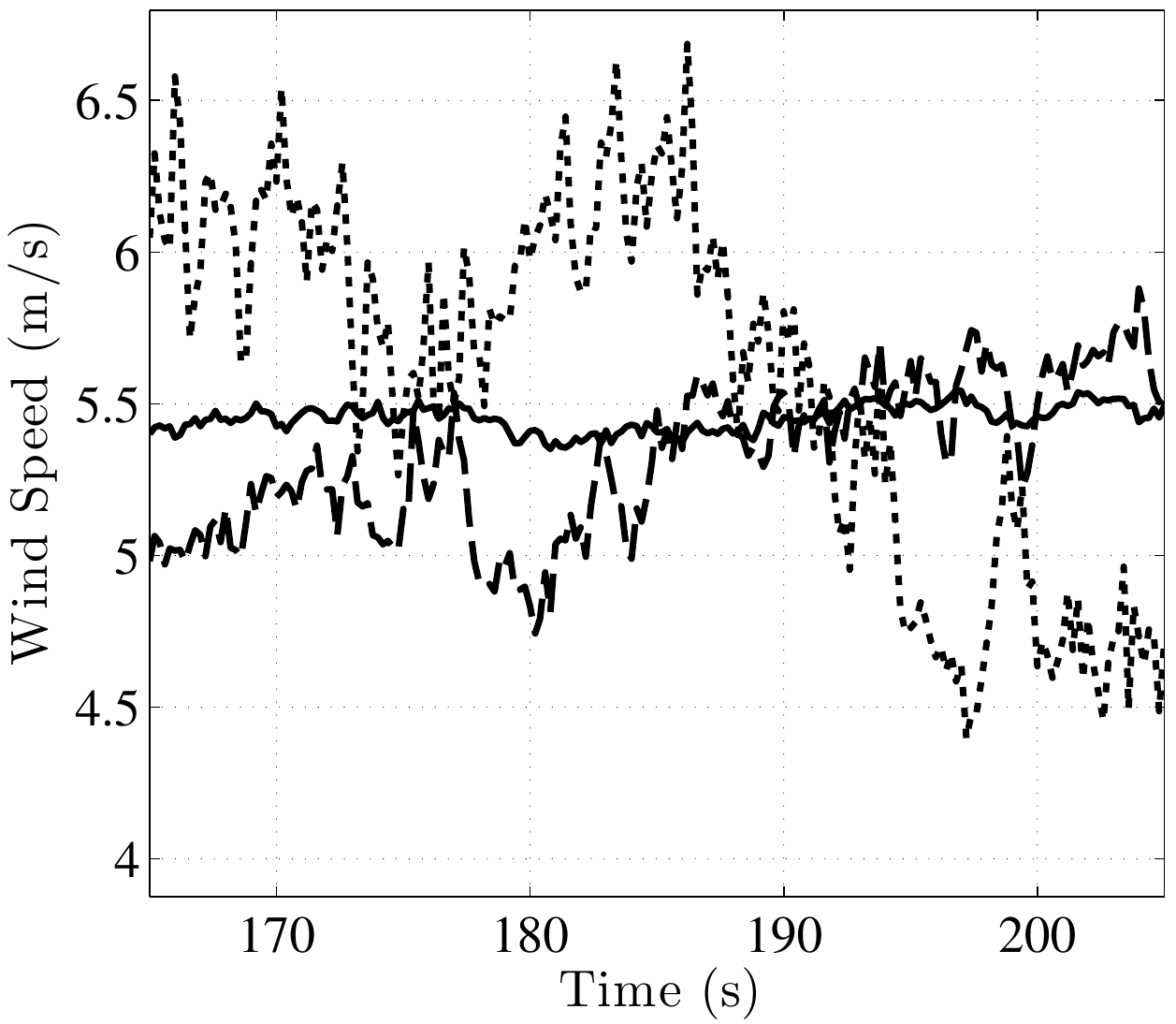}
  \caption{Simulated turbulent wind speed in longitudinal direction over time for three different turbulence intensities with a average wind speed of $\mps{5.4}$, $\perc{1}$ (solid), $\perc{5}$ (dashed), and $\perc{10}$ (dotted).}
  \label{fig:WindFlowTurb}
 \end{center}
\end{figure}

Turbulences are a very complex phenomena, for which a theoretical analysis is difficult to carry out. On the other hand, there exist state-of-the-art turbulence models readily available in public toolboxes, such as TurbSim \cite{TurbSim13}. These can be used to study the effects of turbulences on the system and on the adaptation algorithm via simulations.
The wind fields generated with TurbSim, which is a tool used for wind mill analysis, use a measure for the turbulence strength called intensity $I$, defined as
\begin{IEEEeqnarray*}{rCl}
	I &=& \frac{u'}{U}\, ,
\end{IEEEeqnarray*}
where $U$ is the mean velocity and $u'$ is the root-mean-square of the turbulent velocity fluctuations.
A turbulence intensity of $\perc{10}$ and more is generally considered as strong and $\perc{1}$ to $\perc{5}$ as medium. The generated wind fields provide us with a value of $W_\Delta$ at each sampling time and any point in space. We used TurbSim with the Kaimal power spectrum to generate the turbulence values $W_\Delta$, see \figref{fig:WindFlowTurb} (details about the turbulence model can be found in \cite{TurbSim13}).

\begin{figure}[!htbp]
 \begin{center}
  \includegraphics[trim= 0cm 0cm 0cm 0cm,angle=0,width=\figwidth]{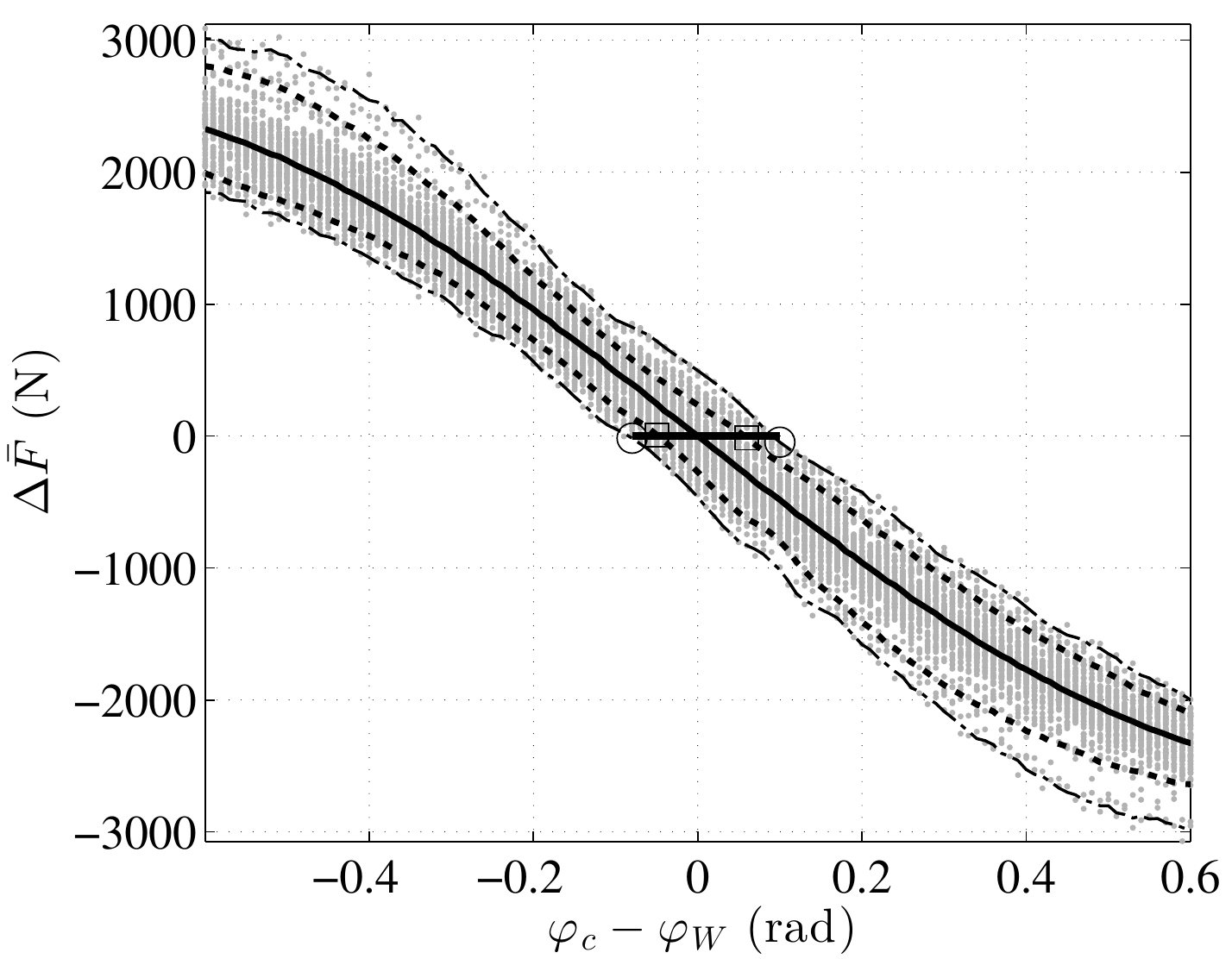}
  \caption{Simulation results. Average traction force differences using the point-mass model for different values of $\phi_c$, with $\theta_c=\arctan{(\sqrt{\alpha})}$, $W_0=\mps{5}$, $I=\perc{5}$. $\Delta\bar{F}$ with no turbulences (solid), $\Delta\bar{F}$ for turbulent wind flow (gray dots), the envelope of $\Delta\bar{F}$ obtained by using $N_{avg}=1$ (dash-dot) and $N_{avg}=5$ (dotted), and $\Delta \phi_c$ (horizontal line) for $N_{avg}=1$ (between circles) and for $N_{avg}=5$ (between squares).}
  \label{fig:PM_DF_TI_Rng}
 \end{center}
\end{figure}

\begin{figure}[!htbp]
 \begin{center}
  \includegraphics[trim= 0cm 0cm 0cm 0cm,angle=0,width=\figwidth]{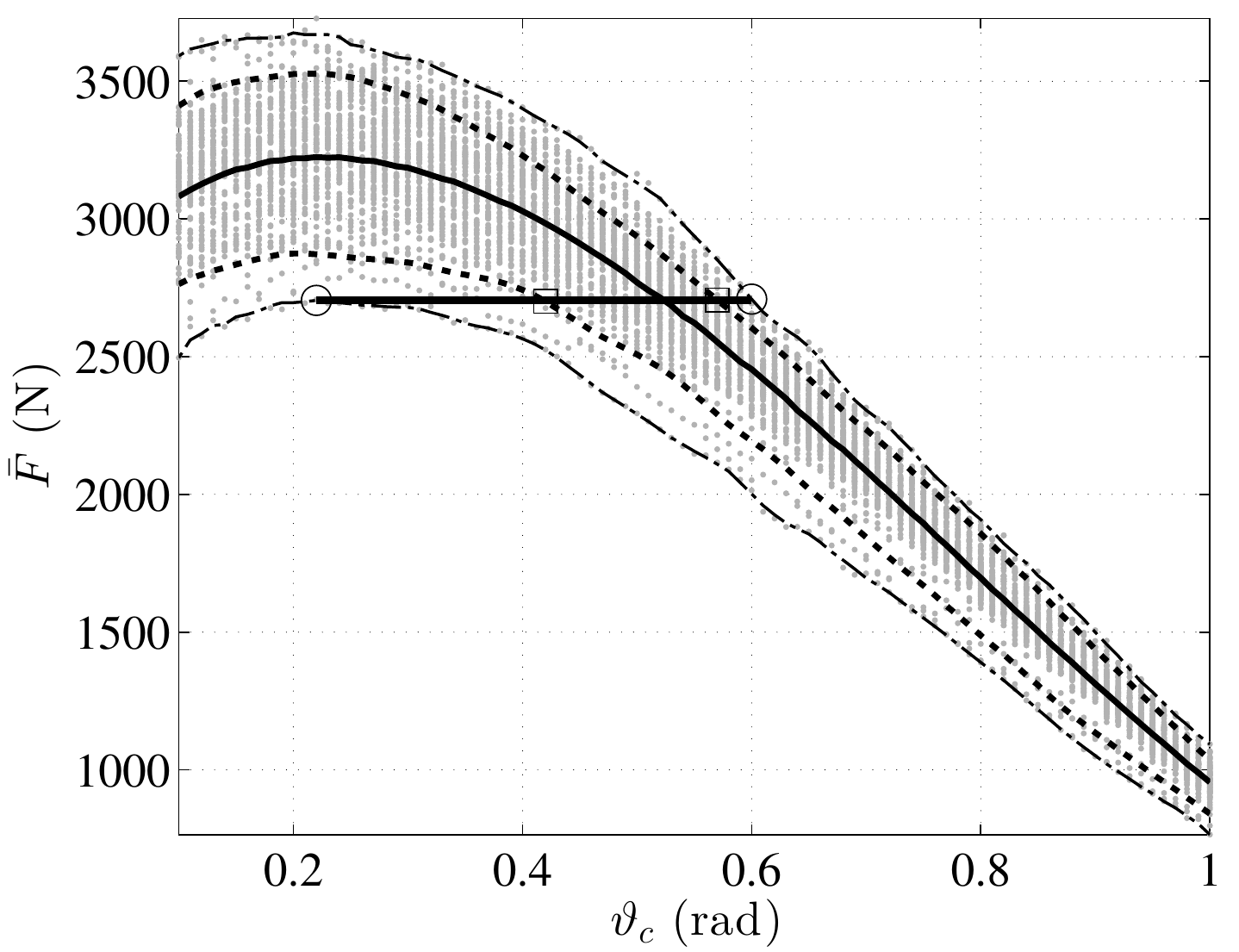}
  \caption{Simulation results. Average traction forces using the point-mass model for different values of $\theta_c$, with $\phi_c=\phi_W$, $W_0=\mps{5}$, $I=\perc{5}$. $\bar{F}$ with no turbulences (solid), $\bar{F}$ for turbulent wind flow (gray dots), the envelope of $\bar{F}$ obtained by using $N_{avg}=1$ (dash-dot) and $N_{avg}=5$ (dotted), and example of interval $\Delta\theta_c$ where the same average traction force can be experienced (horizontal line) with $N_{avg}=1$ (between circles) and with $N_{avg}=5$ (between squares).}
  \label{fig:PM_F_TI_Rng}
 \end{center}
\end{figure}

Due to the turbulent wind, the traction forces experienced during flight will unlikely be equal to their nominal values $\bar{F}$ and $\Delta\bar{F}$, rather they will
lie in an interval around such nominal values, see Figs.~\ref{fig:PM_DF_TI_Rng} and \ref{fig:PM_F_TI_Rng}.

Regarding the $\phi_c$ alignment, i.e. seeking the optimal azimuthal position of the flown path, the presence of turbulence gives rise to a range, $\Delta\phi_c$, of $\phi_c$ values for which a measure of $\Delta\bar{F}=0$ is possible: 
\begin{IEEEeqnarray}{rCl}\label{eqn:DeltaPhiC}
 \Delta\phi_c &=& \max{\left(\phi_c|\Delta\bar{F}=0\right)} - \min{\left(\phi_c|\Delta\bar{F}=0\right)}\, ,\nonumber
\end{IEEEeqnarray}
Thus $\Delta\bar{F}=0$ does not imply that the average loop location is aligned with the nominal wind, i.e. $\phi_c=\phi_W$. Within these azimuthal average positions, the optimization algorithm might make a step in the wrong direction, see \figref{fig:PM_DF_TI_Rng}.
Estimating $\Delta\phi_c$ is not straightforward, since it depends on the current wind situation at the the wing's position. 
However, we can reduce the size of $\Delta\phi_c$ at the expense of convergence speed.
In particular, an intuitive idea to increase the robustness of the approach against turbulences is to use averaged quantities over more than a single flown loop.
This means, instead of comparing the values of a single loop, $N_{avg}>1$ loops are measured before the average values of $\Delta\bar{F}$ and $\bar{F}$ are calculated.
With this approach, equations \eqref{eqn:avgF_k} and \eqref{eqn:avgDF_k} become:
\begin{IEEEeqnarray*}{rCl}
 \bar{F}(\Theta,\phi_W)       &=& \frac{1}{N_{avg}N}\sum_{k=1}^{N_{avg}}{\sum_{k=1}^{N}{F(k)}}\\
 \Delta\bar{F}(\Theta,\phi_W) &=& \frac{1}{N_{avg}}\sum_{k=1}^{N_{avg}}{\bar{F}_{L} -\bar{F}_{R}}\, ,
\end{IEEEeqnarray*}
where $\bar{F}_L$ and $\bar{F}_R$ are from \eqref{eqn:HLavgTF_k}.

This modification can easily be integrated into Algorithm~\ref{algo:BasicMethdod}, by changing the first \textit{if} statement on line \ref{algoLine:firstIF}, which then makes sure that $N_{avg}$ loops are flown instead of just one before the average forces are calculated.

In \figref{fig:PM_DF_TI_Rng}, the values of $\Delta\bar{F}$ as a function of $\phi_c$ in a turbulent wind flow are shown,
for the case $N_{avg}=1$ (gray dots). The same figure shows the envelope of $\Delta\overline{F}$ obtained with $N_{avg}=5$. It can be noted that in this case the use 
of $N_{avg}=5$ decreases $\Delta\phi_c$ by almost a factor of two.

\begin{figure}[!htbp]
 \begin{center}
  \includegraphics[trim= 0cm 0cm 0cm 0cm,angle=0,width=\figwidth]{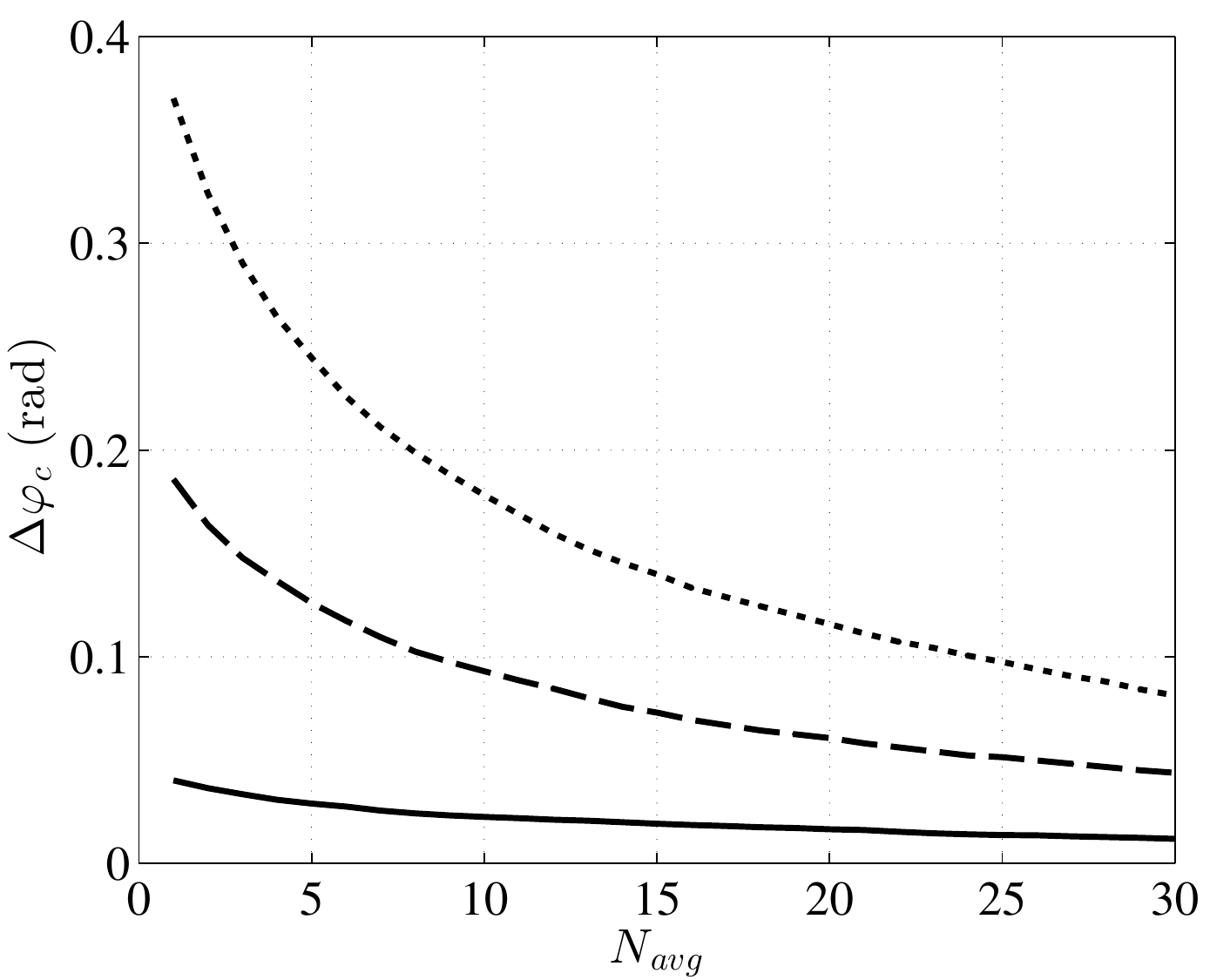}
  \caption{Simulations results. Value of $\Delta\phi_c$ with different turbulence intensities for different values of $N_{avg}$ obtained using the point-mass model. Three different turbulence intensities are shown: $\perc{1}$ (solid), $\perc{5}$ (dashed), $\perc{10}$ (dotted). }
  \label{fig:dPHstpvsNavg}
 \end{center}
\end{figure}

An example of the effect that $N_{avg}$ has on $\Delta\phi_c$ 
for different turbulence intensities is shown in \figref{fig:dPHstpvsNavg}.
For this analysis, $100$ different turbulent wind fields were generated. The point-mass model controlled by controller $K$ was then simulated for $\seconds{300}$. 
The collected data was used to estimate the range $\Delta\phi_c$.
It can be seen that $N_{avg}=5$ gives already a good improvement for strong turbulences. 

Similarly, for the $\theta_c$ alignment there exists a range, $\Delta\theta_c$, of $\theta_c$ values for which a measure of the same $\bar{F}$ value is possible.
Due to the shape of the function $\bar{F}$, this range gets larger closer to the optimal $\theta_c$ value for a given turbulence intensity.
This could lead to steps of the adaptation algorithm away from the optimal elevation of the nominal traction force. Also here, the envelope of $\bar{F}$ is decreased by using $N_{avg}>1$ and thus reducing the sensitivity of the approach with respect to turbulence.
In \figref{fig:PM_F_TI_Rng}, the values of $\bar{F}$ for different values of $\theta_c$ in a turbulent wind flow are shown, together with the effect of using $N_{avg}=5$.
It can be noted that in this case the use of $N_{avg}=5$ decreases $\Delta\theta_c$ by more than a factor of two.

\subsection{Discussion}
We showed that errors in the line angle and force sensors do not impair the performance of our adaptation algorithm.
For the line angle sensor errors, we only considered an additive error with zero mean. Systematic constant errors, such as misalignment of the sensors, do not affect the performance of the algorithm since they would introduce an offset in the force-position curves without altering their qualitative shape, which is exploited by our approach.
For the force sensor errors, we considered a constant multiplicative error and an additive error. Again, these errors do not affect the adaptation algorithm, since the qualitative system behavior is unaffected.
Note that we assumed in both cases a fast sampling rate such that errors from high frequency noise get averaged out over the course of one flown (half-) path. To this end, in our experience a sampling rate of $\Hz{50}$ is sufficiently large to make errors on the computation of the average traction forces negligible.

As a last point, we analyzed the effect of turbulent wind on the adaptation algorithm using a dynamical point-mass model in simulation where the turbulent wind was generated with TurbSim.
We showed that turbulences can cause the algorithm to make steps in the wrong direction around the optimal average location. The range of positions where this can happen increases with the turbulence intensity $I$, but it can be reduced in size by using averaged traction forces over multiple flown loops. Additionally, the stopping criterion $\Delta\bar{F}_{min}$ for the azimuthal position adaptation can be used as a tuning parameter to reduce steps in the wrong direction.

\section{Numerical simulations and experimental results}
\label{sec:results}
We tested the adaptation approach in simulation using the same point-mass dynamical model of the system as in \cite{CaFM09c} and the controller presented in \cite{FaZg13}. The results indicate that the approach is able to tune in real-time the underlying controller $K$ in order to follow a changing wind direction and adapt the paths' average elevation according to the (unknown) wind profile. The main parameters of the model are listed in Table~\ref{ctab:parameters}.

\footnotesize
\ctable[
    caption = Point-Mass Model Parameters used\\ for the numerical simulations,
    cap     = Parameters,
    label   = ctab:parameters,
    pos     = btp, 
    maxwidth = \linewidth,
    captionskip = -3ex
    ]    
    {rclrclrcl}{}
{\FL
$A$       & = & \Meter{9} & $m$       & = & \kg{2.45}     & $r$       & = & \Meter{30}  \NN
$n_l$     & = & $3$       & $d_l$     & = & \Meter{0.003} &           &   &             \NN
$C_L$     & = & $0.8$     & $C_D$     & = & $0.134$       & $C_{D,l}$ & = & $1.2$       \NN
$W_0$     & = & \mps{5}   & $Z_0$     & = & \Meter{4}     & $\alpha$  & = & $0.1$       \LL
}
\normalsize

\begin{figure}[!tbhp]
  \begin{center}
    \includegraphics[trim= 0cm 0cm 0cm 0cm,angle=0,width=\figwidth]{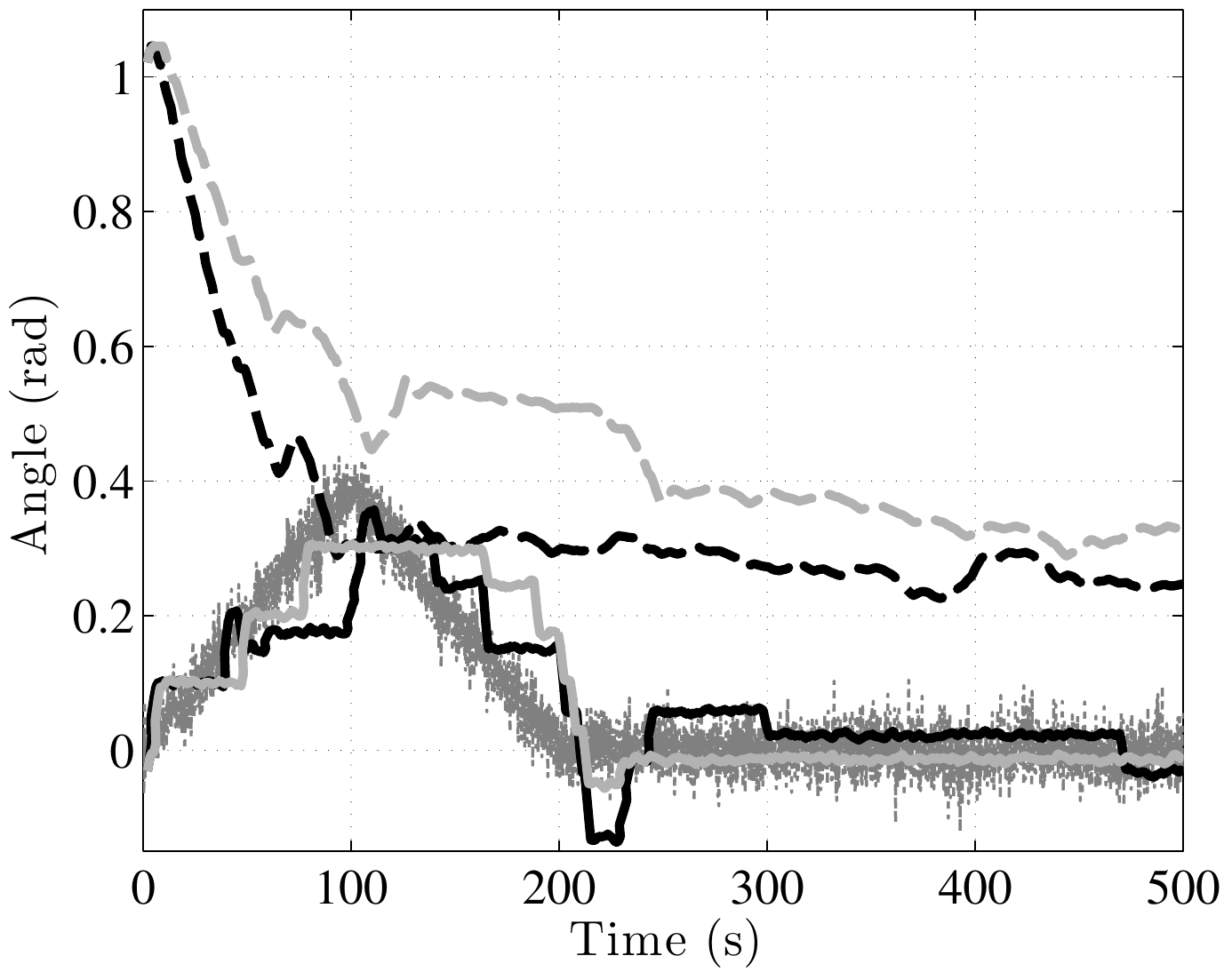}
    \caption{Simulation results obtained by applying the proposed algorithm on the point-mass model with turbulent wind with intensity $\perc{5}$ with $N_{avg}=3$ (black) and $N_{avg}=5$ (light gray). The solid and dashed lines represent the average $\phi$ and $\theta$ positions of the path, $\phi_c$ and $\theta_c$, respectively. The gray dotted line shows the true, turbulent wind direction.}
    \label{fig:THcPHcAdaptation_Sim_Turb}
  \end{center}
\end{figure}

According to simulations, the approach performs well in a turbulent wind field with an appropriate choice of $N_{avg}$. A plot with the time course of the average location of the path and the wind direction in such conditions with $N_{avg}=3$ and $N_{avg}=5$ can be seen in \figref{fig:THcPHcAdaptation_Sim_Turb}.

Additionally, simulation results show that wider loops perform better in turbulent wind situations. This is due to the fact that each half-loop takes longer to complete and thus turbulences get averaged out more than on shorter paths in time. This goes along the same direction as increasing $N_{avg}$.

\begin{figure}[!tbhp]
 \begin{center}
  \includegraphics[trim= 0cm 0cm 0cm 0cm,angle=0,width=\figwidth]{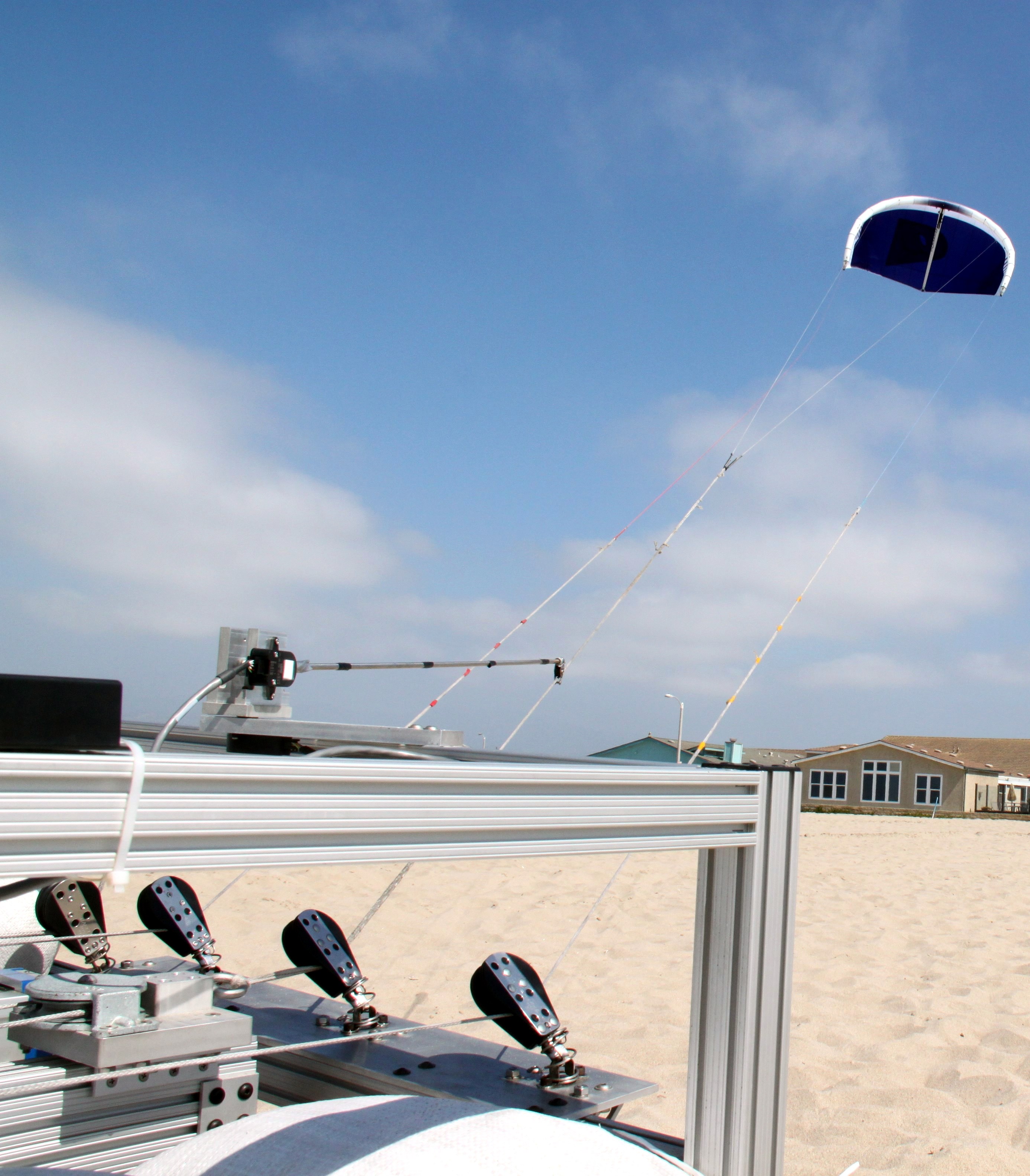}
  \caption{Small-scale prototype built at the University of California, Santa Barbara, to study the control of tethered wings for airborne wind energy.}
  \label{F:proto}
 \end{center}
\end{figure}

Experimental test flights using the presented algorithm have also been carried out on a small scale prototype (shown in \figref{F:proto}), built at UC Santa Barbara, with promising results. The prototype used two different three-line, inflatable kites with a constant tether length of $r=\Meter{30}$. The employed power kites were Airush One $\sqm{6}$ and $\sqm{9}$ kites. 
Due to the short lines, a measurement of $\phi_W$ with good accuracy was possible with an anemometer approximately $\Meter{4}$ above the ground. The algorithm was set to use $N_{avg}=3$. For more details on the test setup, see \cite{FaZg13} and \cite{FaMa13}.

\begin{figure}[!tbhp]
  \begin{center}
    \includegraphics[trim= 0cm 0cm 0cm 0cm,angle=0,width=\figwidth]{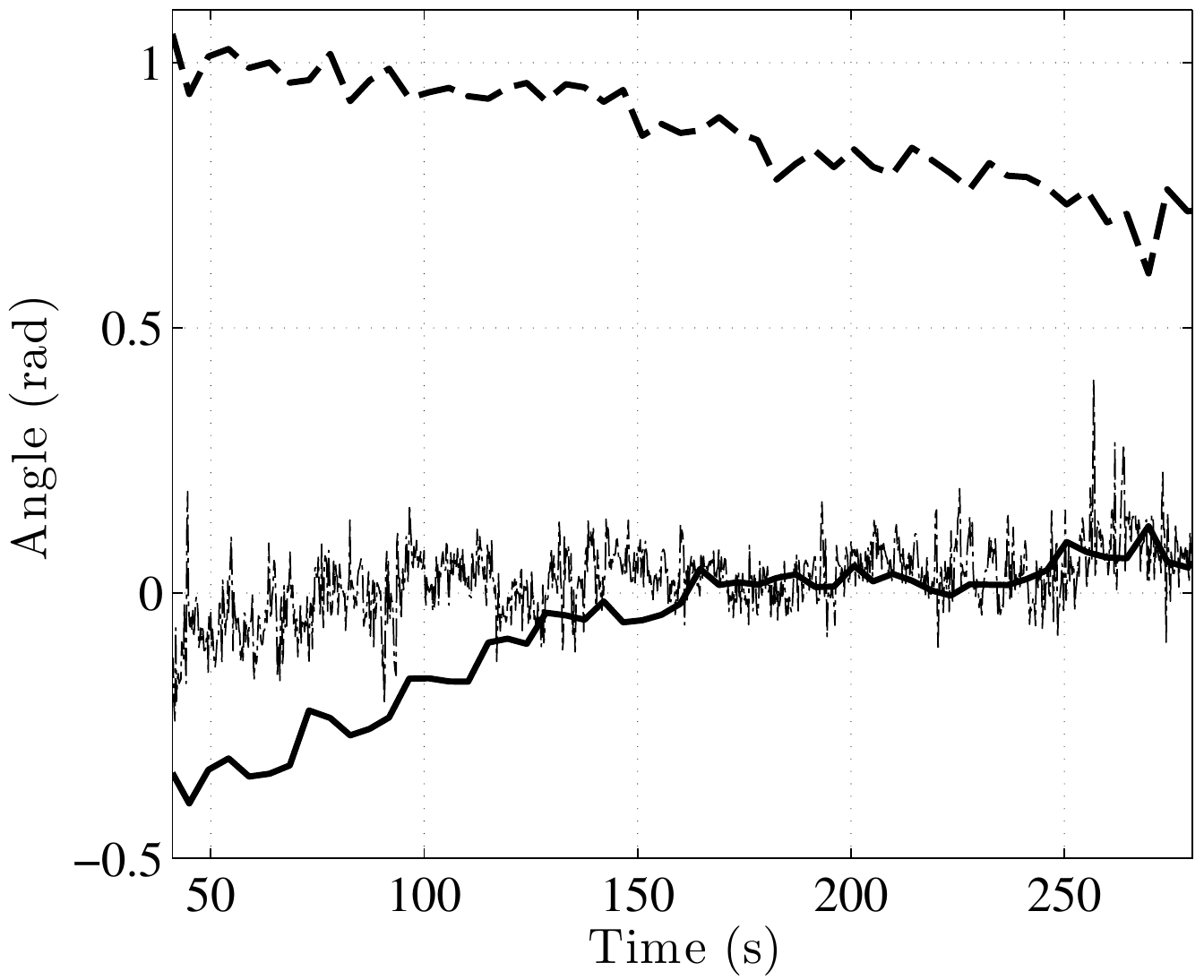}
    \caption{Experimental test results using a small-scale prototype with a $\sqm{9}$ kite. The $\phi_c$ position (solid) and $\theta_c$ position (dashed) of the paths, and the wind direction $\phi_W$ (dotted) are shown.}
    \label{fig:THcPHcAdaptation_9sqm}
  \end{center}
\end{figure}

\begin{figure}[!tbhp]
  \begin{center}
    \includegraphics[trim= 0cm 0cm 0cm 0cm,angle=0,width=\figwidth]{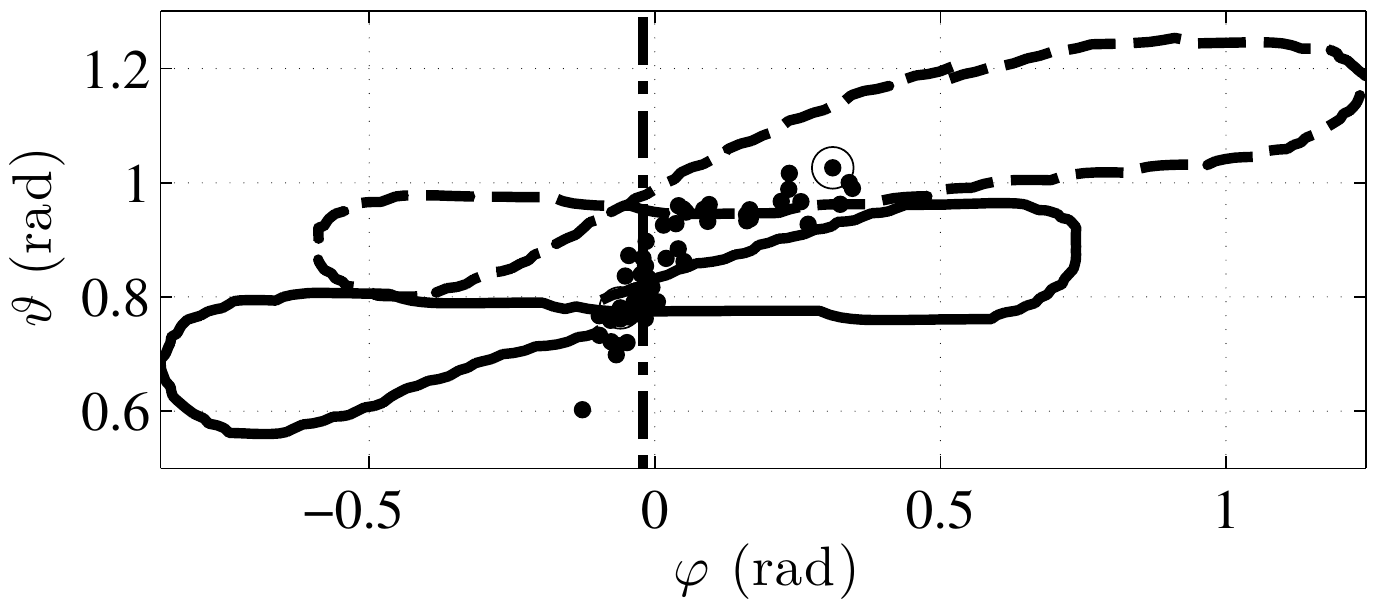}
    \caption{Experimental test results using a small-scale prototype with a $\sqm{9}$ kite. Initial (dashed) and final (solid) paths flown by the wing corresponding to the data shown in Figs.~\ref{fig:THcPHcAdaptation_9sqm}, \ref{fig:THcPHcAdaptation_Force_9sqm}, and \ref{fig:THcPHcAdaptation_WindSpd_9sqm}. The trajectory of $(\phi_c,\theta_c)$ (dotted) and the initial and final $(\phi_c,\theta_c)$ locations (circles) are shown together with the optimal $\phi_c^*$ location (dashed-dotted).}
    \label{fig:THcPHcAdaptation_Loops_9sqm}
  \end{center}
\end{figure}

\begin{figure}[!tbhp]
  \begin{center}
    \includegraphics[trim= 0cm 0cm 0cm 0cm,angle=0,width=\figwidth]{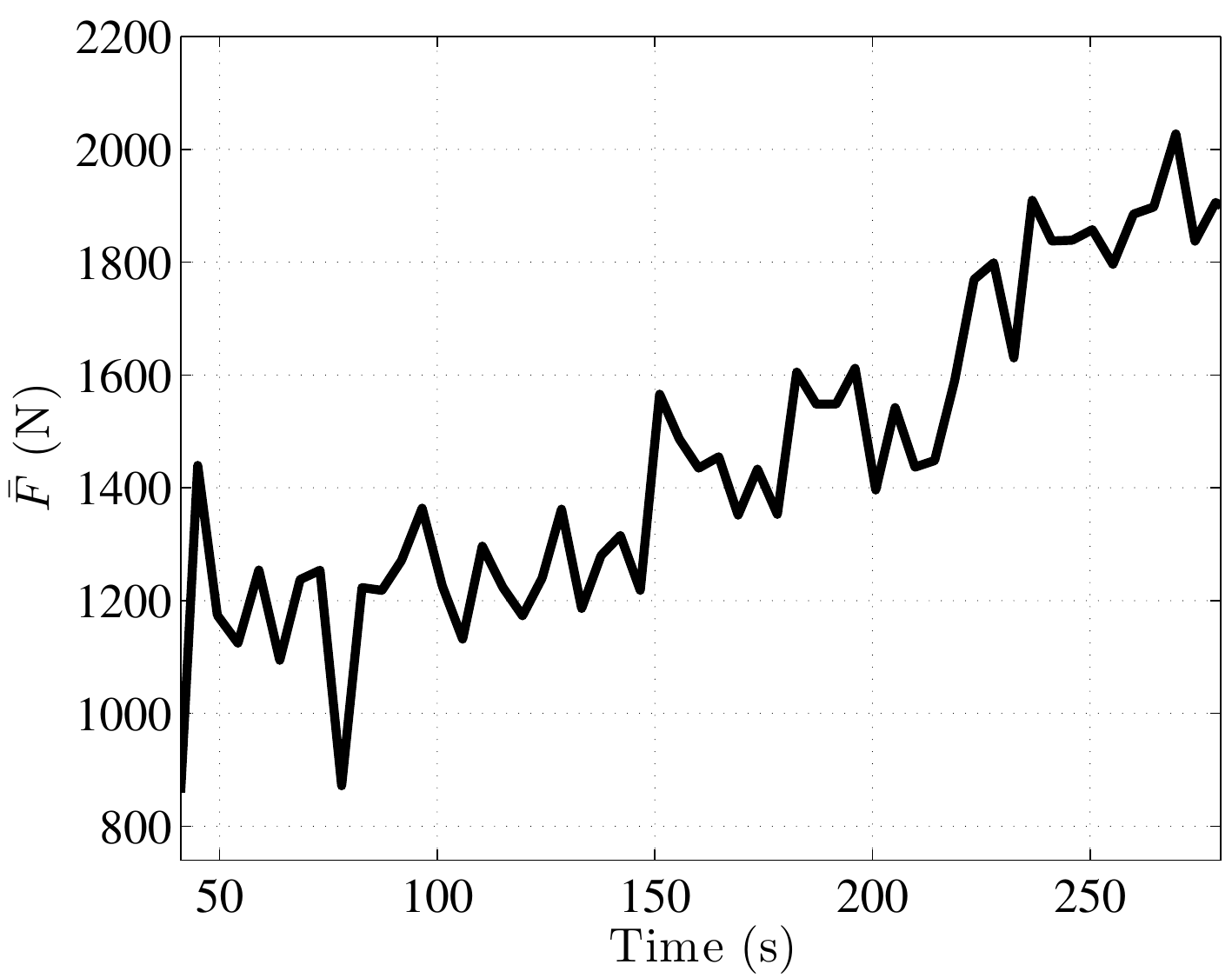}
    \caption{Experimental test results using a small-scale prototype with a $\sqm{9}$ kite. Course of the average traction force $\bar{F}$.}
    \label{fig:THcPHcAdaptation_Force_9sqm}
  \end{center}
\end{figure}

\begin{figure}[!tbhp]
  \begin{center}
    \includegraphics[trim= 0cm 0cm 0cm 0cm,angle=0,width=\figwidth]{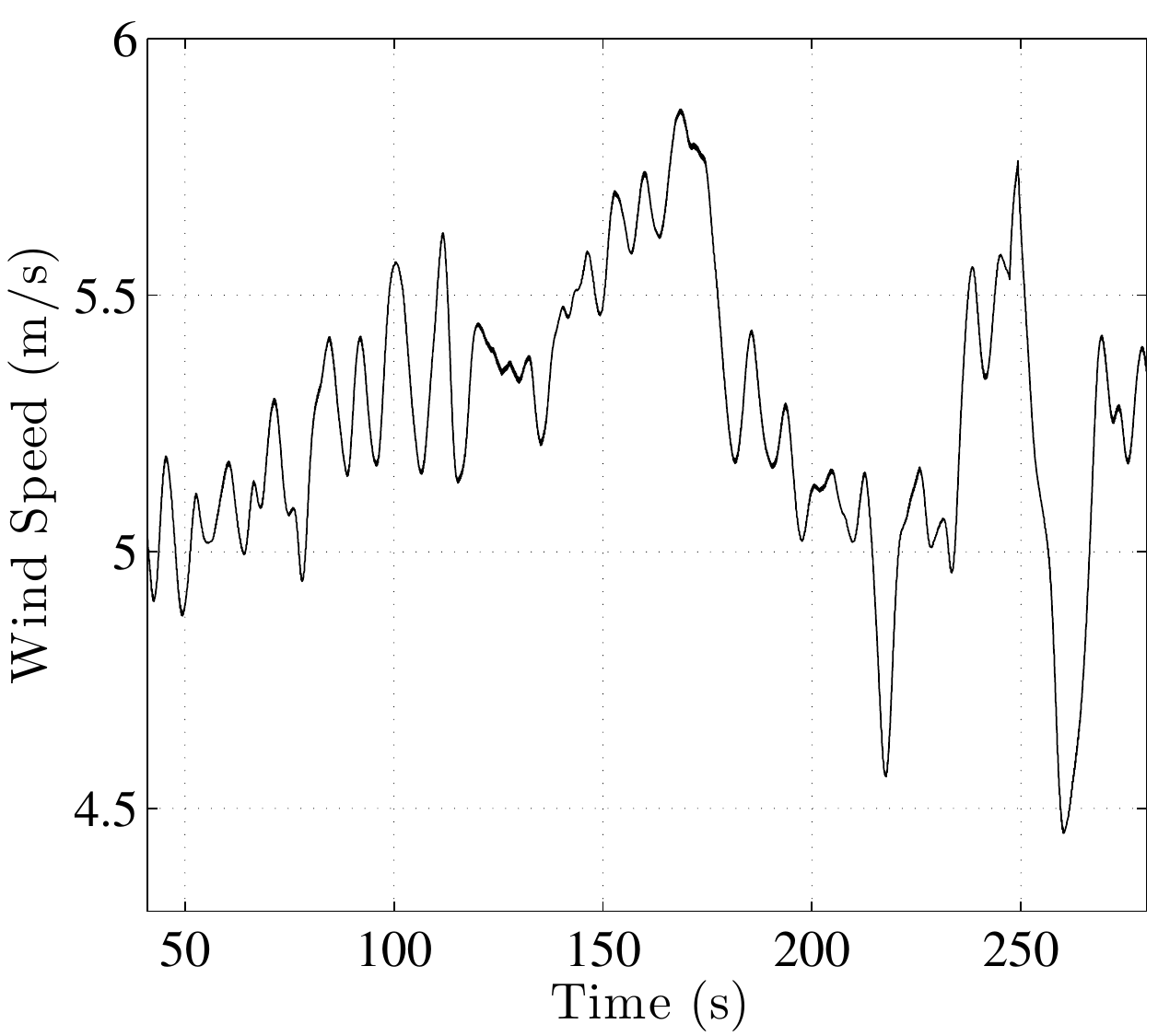}
    \caption{Experimental test results using a small-scale prototype with a $\sqm{9}$ kite. Course of wind speed measured roughly \Meter{4} above the ground. The average wind speed was \mps{5.3}.}
    \label{fig:THcPHcAdaptation_WindSpd_9sqm}
  \end{center}
\end{figure}

A test flight with the $\sqm{9}$ kite is reported in Figs.~\ref{fig:THcPHcAdaptation_9sqm}-\ref{fig:THcPHcAdaptation_WindSpd_9sqm}. The underlying controller was initialized to fly a path with a misalignment of roughly $\degr{20}$ from the wind direction $\phi_W$ and with a overly high elevation. The algorithm was able to correct the misalignment with the wind direction in the first $\seconds{150}$, i.e. roughly 30 flown loops, and then to adapt the azimuthal position according to wind direction changes while improving the average elevation, see \figref{fig:THcPHcAdaptation_9sqm}. The initial and final paths of the wing with the measured trajectory $(\phi_c,\theta_c)$ can be seen in \figref{fig:THcPHcAdaptation_Loops_9sqm}. Note that although the paths seem not to differ much in position, yet the force increase is significant, see \figref{fig:THcPHcAdaptation_Force_9sqm}, as expected from the sensitivity analysis presented in section \ref{sec:CWTF}. The corresponding wind speed during the test flight can be seen in \figref{fig:THcPHcAdaptation_WindSpd_9sqm}, with an average value of $\mps{5.3}$. 
Note that due to the alignment with the wind the loop becomes more symmetric, thus indicating that measures of elapsed time or speed of the wing in a half path could also potentially be used, instead or in addition to the force, to detect a misalignment with the wind direction.

\begin{figure}[!tbhp]
  \begin{center}
    \includegraphics[trim= 0cm 0cm 0cm 0cm,angle=0,width=\figwidth]{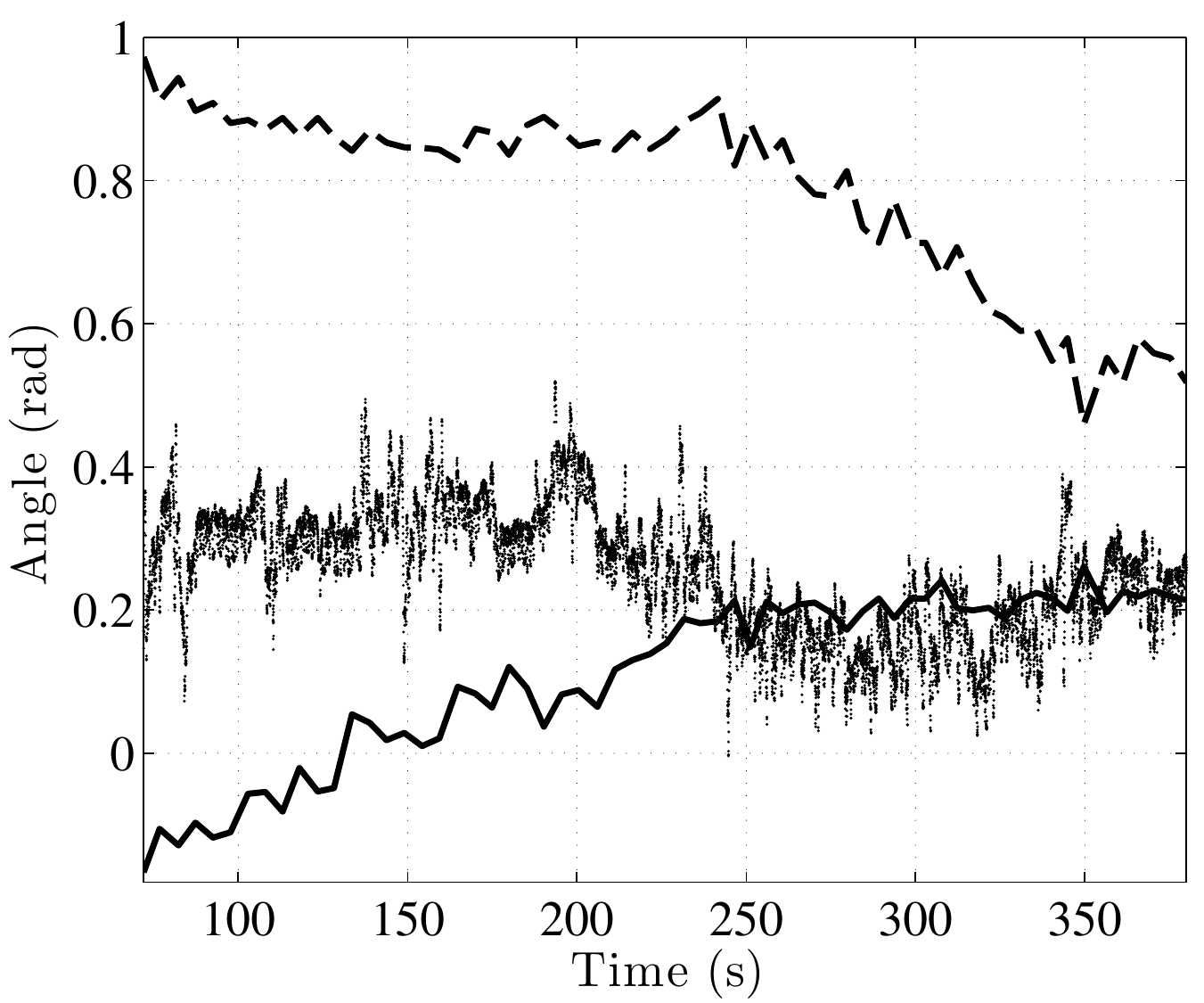}
    \caption{Experimental test results using a small-scale prototype with a $\sqm{6}$ kite. The $\phi_c$ position (solid) and $\theta_c$ position (dashed) of the paths, and the wind direction $\phi_W$ (dotted) are shown.}
    \label{fig:THcPHcAdaptation_6sqm}
  \end{center}
\end{figure}

\begin{figure}[!tbhp]
  \begin{center}
    \includegraphics[trim= 0cm 0cm 0cm 0cm,angle=0,width=\figwidth]{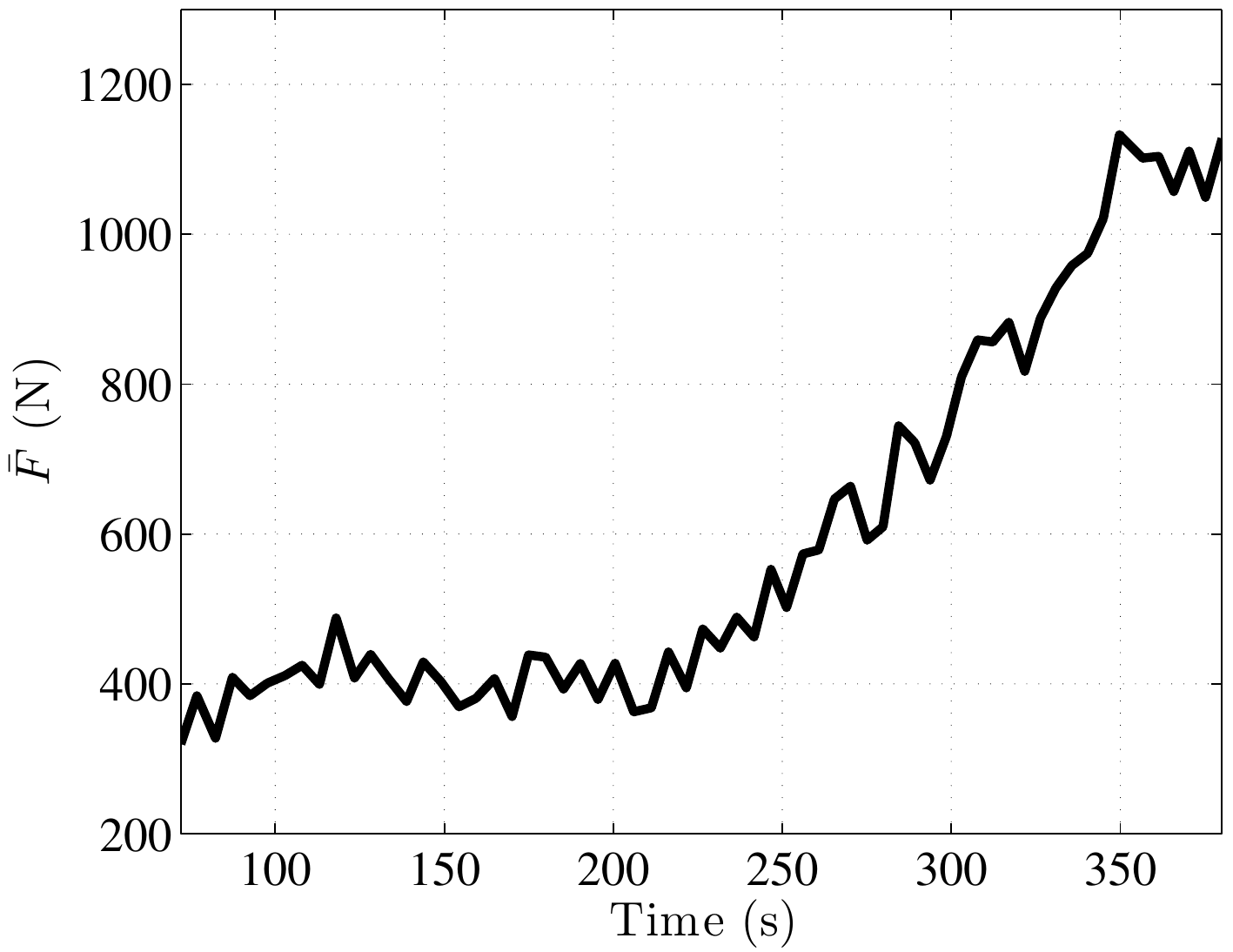}
    \caption{Experimental test results using a small-scale prototype with a $\sqm{6}$ kite. Course of the average traction force $\bar{F}$.}
    \label{fig:THcPHcAdaptation_Force_6sqm}
  \end{center}
\end{figure}

\begin{figure}[!tbhp]
  \begin{center}
    \includegraphics[trim= 0cm 0cm 0cm 0cm,angle=0,width=\figwidth]{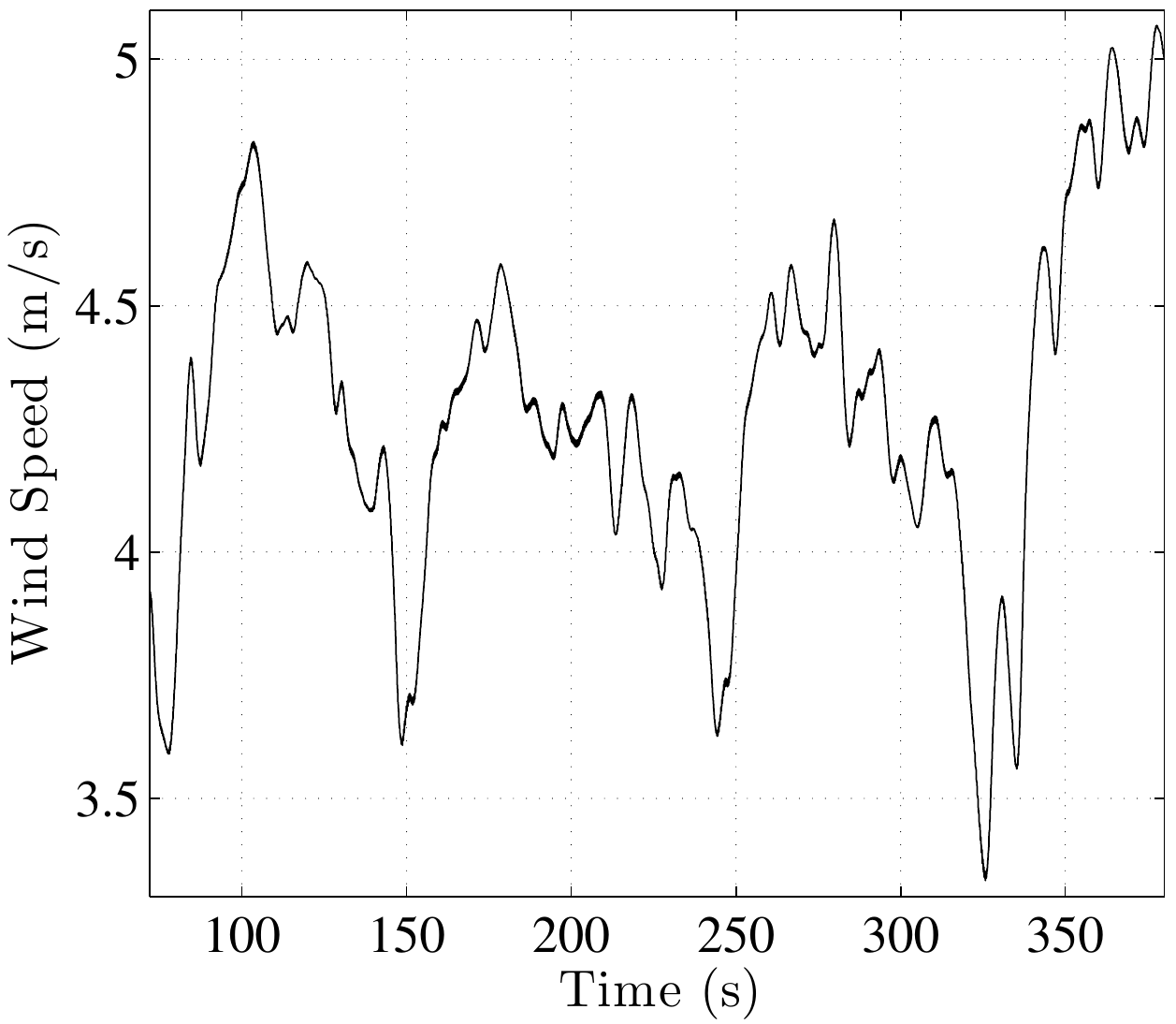}
    \caption{Experimental test results using a small-scale prototype with a $\sqm{6}$ kite. Course of wind speed measured roughly \Meter{4} above the ground. The average wind speed was \mps{4.3}.}
    \label{fig:THcPHcAdaptation_WindSpd_6sqm}
  \end{center}
\end{figure}

\begin{figure}[!tbhp]
  \begin{center}
    \includegraphics[trim= 0cm 0cm 0cm 0cm,angle=0,width=\figwidth]{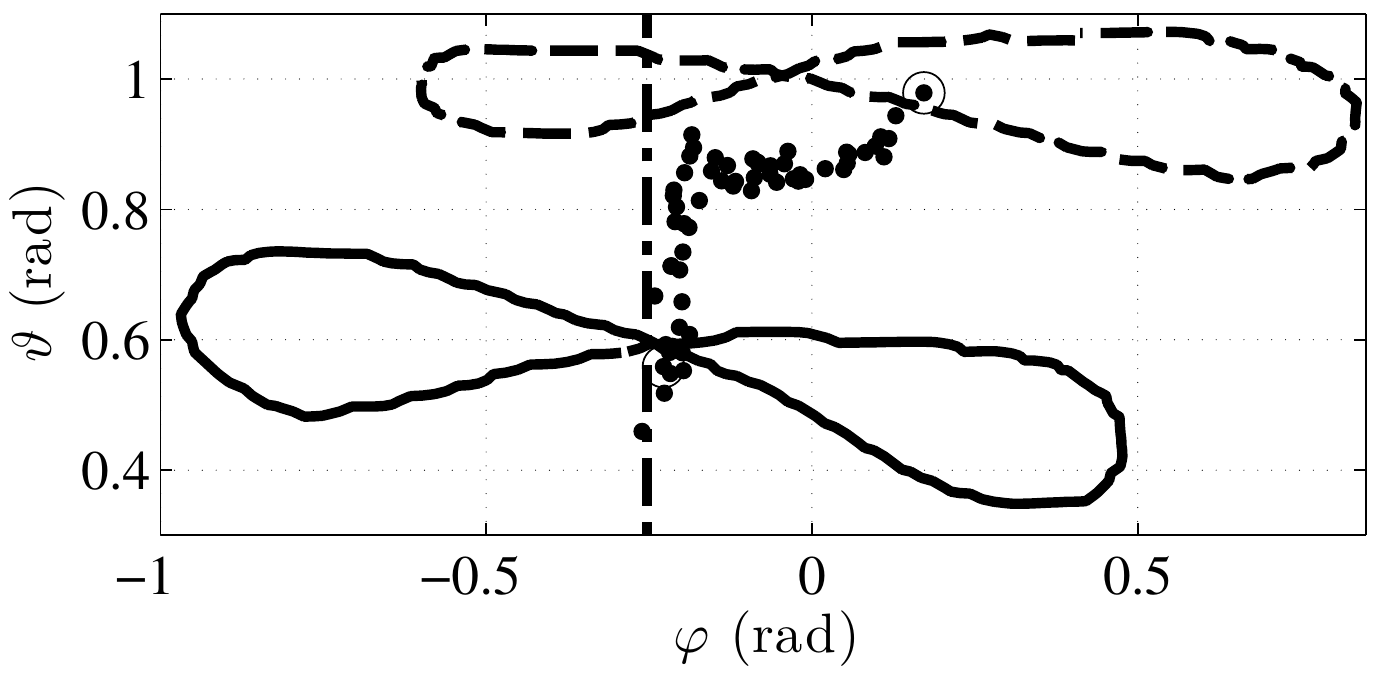}
    \caption{Experimental test results using a small-scale prototype with a $\sqm{6}$ kite. Initial (dashed) and final (solid) paths flown by the wing corresponding to the data shown in Figs.~\ref{fig:THcPHcAdaptation_6sqm}-\ref{fig:THcPHcAdaptation_WindSpd_6sqm}. The trajectory of $(\phi_c,\theta_c)$ (dotted) and the initial and final $(\phi_c,\theta_c)$ locations (circles) are shown together with the optimal $\phi_c^*$ location (dashed-dotted).}
    \label{fig:THcPHcAdaptation_Loops_6sqm}
  \end{center}
\end{figure}

A test flight with the $\sqm{6}$ kite, where it was initially commanded to fly a path with a misalignment of roughly $\degr{25}$ from the wind direction $\phi_W$ and with a overly high elevation, is reported in Figs.~\ref{fig:THcPHcAdaptation_6sqm}-\ref{fig:THcPHcAdaptation_Loops_6sqm}. Also in this case, the algorithm is able to first correct the misalignment with the wind and then to improve the traction force by changing $\theta_c$. A short movie of the test with the adaptive algorithm and this kite is also available online \cite{Wing_movieAdaptive}. 
In \figref{fig:THcPHcAdaptation_6sqm}, the time courses of the wind direction $\phi_W$ and of the average position $\phi_c$ and $\theta_c$ of the path, modified in real-time by the proposed algorithm, can be seen. In \figref{fig:THcPHcAdaptation_Force_6sqm}, the corresponding average traction force for each full path is shown. It can be noted that the average force increases significantly thanks to the adaptive approach.
The time course of the wind speed magnitude is shown in \figref{fig:THcPHcAdaptation_WindSpd_6sqm}. 
Finally, \figref{fig:THcPHcAdaptation_Loops_6sqm} shows two measured flown paths, at the beginning and at the end of one test, together with the optimal location in terms of average angle $\phi_c$ and with the measured trajectory of ($\phi_c,\theta_c$).

\section{Conclusion an future work}
We presented an analysis of the average traction force generated by a tethered wing and, based on the results of such analysis, we proposed an algorithm to adapt and optimize in real-time the average position of the flown path without exact knowledge of the wind direction and profile.
The algorithm is not dependent on the system configuration, e.g. number of lines or position of the generator, and 
it can be used as an extension of any working controller for a tethered wing, provided that the controller is able to control the wing in order to fly on a symmetric horizontal path and to attain a reference position in terms of average location of the path in the wind window.
We tested the approach both with numerical simulations and real-world experiments, showing good performance in adapting and optimizing the system's operation in the presence of unknown and changing wind conditions with turbulences.

\appendix[Detailed Adaptation Algorithm]
\small
The outline of the Algorithm~\ref{algo:BasicMethdodDetailed} below is a more detailed version of Algorithm~\ref{algo:BasicMethdod}. Note that this algorithm uses $N_{avg}$ loops to calculate the average forces.
For path-related variables, we use $i$ as the index standing for the last $N_{avg}$ flown full paths, e.g. $\bar{F}(i)$ is the average traction force of the last $N_{avg}$ paths and $\theta_c(i)$ the average $\theta$ position of the last $N_{avg}$ paths. The employed coordinate search method uses the step sizes $\delta_\phi$ and $\delta_\theta$ for the adaptation of the $\phi_c$ and $\theta_c$ directions. Both step sizes have a defined minimal and maximal value, denoted by a subscript $min$ or $max$. At each change in $\phi_c$ or $\theta_c$, the related step size is adapted with a scaling factor $c>1$ (if the step direction is unchanged) or $1/c$ (if the step direction changes).  

\begin{algorithm}[htbp]
  \label{algo:BasicMethdodDetailed}
  \caption{Optimization/Adaptation - Detailed}
  \DontPrintSemicolon
  \While{ true }{
    \If{ $N_{avg}$ complete loops flown }{
      \eIf{ $|\Delta\bar{F}(i)| > \Delta\bar{F}_{min}$ }{
	\eIf{$\Delta\bar{F}(i) > 0$}{
	  \eIf{$\Delta\bar{F}(i-1) > 0$}{
	    $\delta_\phi = \min\{\delta_{\phi,max},c\:\delta_\phi\}$\;
	  }{
	    $\delta_\phi = \max\{\delta_{\phi,min},\frac{1}{c}\delta_\phi\}$\;
	  }
	  $\phi_c(i+1) = \phi_c(i) + \delta_\phi$\;
	}(\tcc*[f]{$\Delta\bar{F}(i) < 0$}){
	  \eIf{$\Delta\bar{F}(i-1) < 0$}{
	    $\delta_\phi = \min\{\delta_{\phi,max},c\delta_\phi\}$\;
	  }{
	    $\delta_\phi = \max\{\delta_{\phi,min},\frac{1}{c}\delta_\phi\}$\;
	  }
	  $\phi_c(i+1) = \phi_c(i) - \delta_\phi$\;
	}
      }{
      \eIf{$\bar{F}(i-1) < \bar{F}(i)$}{
	  $\delta_\theta = \min\{\delta_{\theta,max},c\:\delta_\theta\}$\;
	  \eIf{$\theta_c(i-1) > \theta_c(i)$ }{
	    $\theta_c(i+1) = \theta_c(i) - \delta_\theta$\;
	  }{
	    $\theta_c(i+1) = \theta_c(i) + \delta_\theta$\;
	  }
      }(\tcc*[f]{$\bar{F}(i-1) \geq \bar{F}(i)$}){
	  $\delta_\theta = \max\{\delta_{\theta,min},\frac{1}{c}\delta_\theta\}$\;
	  \eIf{$\theta_c(i-1) > \theta_c(i)$}{
	    $\theta_c(i+1) = \theta_c(i) + \delta_\theta$\;
	  }{
	    $\theta_c(i+1) = \theta_c(i) - \delta_\theta$\;
	  }
      }
      }
    }
  }
\end{algorithm}

\bibliographystyle{IEEEtran}

\end{document}